\documentclass[pdflatex,sn-mathphys-num]{sn-jnl}

\usepackage{graphicx}%
\usepackage{multirow}%
\usepackage{amsmath,amssymb,amsfonts}%
\usepackage{amsthm}%
\usepackage{mathrsfs}%
\usepackage[title]{appendix}%
\usepackage{xcolor}%
\usepackage{textcomp}%
\usepackage{manyfoot}%
\usepackage{booktabs}%
\usepackage{algorithm}%
\usepackage{algorithmicx}%
\usepackage{algpseudocode}%
\usepackage{listings}%
\usepackage[subrefformat=parens]{subcaption}
\usepackage{bm}

\usepackage{xcolor}
\usepackage{soul}

\theoremstyle{thmstyleone}%
\theoremstyle{thmstyletwo}%

\theoremstyle{thmstylethree}%

\begin{document}

\title[Article Title]{Predicting trigger in streak breakdown via Picard-Chebyshev approach in plane-Poiseuille flow}


\author*[1]{Nobutaka Taniguchi}\email{nobutaka.taniguchi.c1@tohoku.ac.jp}

\author[1]{Aiko Yakeno}\email{aiko.yakeno@tohoku.ac.jp}
\affil*[1]{\orgdiv{Institute of Fluid Science}, \orgname{Tohoku University}, \orgaddress{\street{2-1-1 Katahira}, \city{Sendai}, \postcode{9808577}, \state{Miyagi}, \country{Japan}}}


\abstract{
The evolution of finite-amplitude disturbances is critical in the prediction of laminar-to-turbulent transition. To analyze the transient growth of disturbances, Pralits et al. (J. Fluid Mechanics, 2015) proposed a weakly nonlinear optimal perturbation (WNLOP) analysis that accounts for the effect of mean-flow distortion on the fundamental wave. The present study formulates a WNLOP analysis via Picard--Chebyshev iteration for compressible parallel flows to predict streak breakdown. The formulation incorporates the non-modal amplification of disturbances and their nonlinear interactions for polychromatic waves, considering all the possible nonlinear interactions starting from a finite set of initial wavenumber components. The nonlinear interaction is tracked as a monotonically increasing set of wavenumber components; specifically, the first-order approximation recovers the linear disturbance equation, the third-order approximation corresponds to the amplitude equation of weakly nonlinear theory, and further iterations systematically approach DNS results in the weakly nonlinear regime. By applying the proposed formulation to a subcritical plane Poiseuille flow at Mach number $\mathrm{M}=0.3$ and Reynolds number $\mathrm{Re}=3000$, the proposed formulation demonstrates high predictive accuracy in comparison with DNS calculations and succeeds in identifying the optimal perturbation restricted to a planar wave that causes streak breakdown in full nonlinear simulation. The proposed method predicts the nonlinear interaction between the streak and the planar wave, which generates an oblique wave and subsequent harmonic waves characterized by an alternating streamwise vortex structure, whose amplitude determines whether further nonlinear interactions are induced to trigger the turbulent transition. 
}

\keywords{Nonlinear dynamics, transition theory, shear layer}



\maketitle

\section{Introduction}

Non-modal analysis is a standard tool to analyze the stability properties of viscous channel flows. The linear non-modal amplification is attributed to the non-orthogonality of the eigenmodes of the linearized Navier--Stokes (N--S) operator in linear stability theory (LST), which allows the transient growth of disturbances even under linearly subcritical conditions. Reddy and Henningson~\cite{reddy1993energy} reported a detailed investigation on the properties of the linearized N--S operator in plane Poiseuille and Couette flows by identifying the linear optimal perturbation (LOP), including the limit regime of transient growth in relation to an absolutely stable state and the dependency of the growth rate on spatial wavenumber. The lift-up mechanism predicted by the LOP forms streamwise streak structures with locally large amplitudes of streamwise velocity~\cite{trefethen1993hydrodynamic}, which provides a fundamental understanding of the streamwise vortices (SV) and oblique waves (OW) scenarios in Reddy et al.~\cite{reddy1998stability}. 

The effect of nonlinearity in the evolution of disturbances has been extensively studied for decades using both the weakly nonlinear theory and full numerical calculations. The evolution of the amplitude for a single eigenmode is described by the Landau equation (\S 26 in Landau and Lifshitz~\cite{landau2013fluid} and \S 49 in Drazin and Reid~\cite{drazin2004hydrodynamic}) with a natural extension to amplitude equations for multiple modes. In cases of non-modal analysis, the effect of nonlinearity after the transient growth was investigated in the pioneering studies~\cite{trefethen1993hydrodynamic, butler1994nonlinear}. The full consideration of nonlinear effects in disturbance growth has been accomplished by Pringle and Kerswell~\cite{pringle2010using} and Cherubini et al.~\cite{cherubini2010rapid} as a variational formulation for obtaining nonlinear optimal perturbations (NLOPs), later extended to compressible flows~\cite{huang2020variational, taniguchi2026minimal}. In a review article, Kerswell~\cite{kerswell2018nonlinear} classified optimal perturbations based on the degree of nonlinear effects into quasi-linear optimal perturbations (QLOPs) and nonlinear optimal perturbations (NLOPs). They noted that nonlinearity initially suppresses linear amplification while preserving the qualitative structure of LOPs in the case of QLOPs. After exceeding a certain critical value of the initial disturbance magnitude, the evolution changes into a qualitatively different path, characteristic of an NLOP. 

For incompressible boundary layer flow, Cherubini et al.~\cite{cherubini2010rapid} reported that NLOPs showed spatial modulation and localization, and they demonstrated that these structural changes led to transition. The behavior of NLOPs in plane Poiseuille flow was investigated in detail by Farano et al.~\cite{farano2016subcritical}, where the parameter space defined by initial disturbance amplitude and evaluation time can be categorized into LOP, weakly NLOP, hairpin-like NLOP, and highly NLOP to represent the different paths of laminar-to-turbulent transition. It is shown that the spatial localization in each direction incrementally occurs as the degree of nonlinearity increases. 

In spite of these advancements in nonlinear dynamics, the high computational cost associated with determining NLOPs hampers wide applications of these approaches to the laminar-to-turbulent transition process. Recently, Pralits et al.~\cite{pralits2015weakly} proposed a framework of weakly nonlinear optimal perturbation (WNLOP), which incorporates the mean flow deformation effect on the evolution of disturbances in a self-consistent model. Ducimeti{\`e}re and Gallaire~\cite{ducimetiere2025nonmodal} approached the evolution of finite-amplitude disturbances as an amplitude equation of the LOP. However, these formulations are essentially restricted to the evolution of a single-mode with the mean-flow-distortion effect, making it challenging to stably capture higher-order nonlinear couplings among polychromatic waves. 

To overcome these limitations, the present study aims to extend the WNLOP framework into a comprehensive WNLOP analysis for parallel flows, capable of systematically incorporating high-order nonlinear interactions within a weakly nonlinear regime among polychromatic waves via Picard--Chebyshev (P--C) iterations. The proposed formulation tracks a monotonically expanding set of wavenumber components generated from a finite set of initial wavenumber components. Through the application of the present formulation to plane Poiseuille flow at Mach number $\mathrm{M}=0.3$ and Reynolds number $\mathrm{Re}=3000$, we demonstrate that the proposed method contains weakly nonlinear theory as a third-order truncation model and approaches DNS calculations within a weakly nonlinear regime, identifying the existence of a key vortex structure responsible for streak breakdown. This paper is organized as follows: Sec. \ref{sec:Formulation of Picard--Chebyshev (P--C) approximation} presents the formulation of Picard--Chebyshev method for general nonlinear partial differential equations and Sec. \ref{sec:Governing equations and numerical methods} provides the governing equations, amplitude equations formulation in weakly nonlinear theory (WNT) for comparison, and the numerical implementation. Main results are presented in Sec. \ref{sec:Numerical results}, starting with the verification of the model for monochromatic waves in comparison with the WNT model and DNS. The latter part presents the investigation of streak breakdown through the formulated approach.

\section{Formulation of Picard--Chebyshev (P--C) approximation}
\label{sec:Formulation of Picard--Chebyshev (P--C) approximation}

This study seeks an approximate solution of the quadratic nonlinear partial differential equations: 
\begin{equation}
\label{equ:equ_quad_dis_equ}
\partial_t \bm{q}' = L (\bar{\bm{q}}) \bm{q}' + N (\bm{q}', \bm{q}'), 
\end{equation}
where $L$ is a linear operator and $N (\bm{q}', \bm{q}')$ is a quadratic nonlinear operator. Here, $\bm{q}' = \bm{q}' (t, \bm{x})$ and $\bm{x} \in [0, L_x] \times [0, L_y] \times [-1, 1]$ where $L_x, L_y$ are chosen as integer multiples of $2 \pi$ to allow fractional wavenumbers, which is required in the subsequent discussion to consider the nonlinear interactions of two fundamental waves. The function $\bar{\bm{q}} = \bar{\bm{q}} (z)$ is a given vector with sufficient smoothness. We assume the periodicity over $x$ and $y$ with periods of $L_x, L_y$, respectively. Under appropriate functional assumptions, the equation for wavenumber $k$ reduces to 
\begin{equation}
\label{equ:non_dis_wave}
\partial_t \hat{\bm{q}}_k = \hat{L}_k (\bar{\bm{q}}) \hat{\bm{q}}_k + \hat{N}_k (t, \bm{q}'), 
\end{equation}
where $\hat{L}_k$ and $\hat{N}_k$ are for this wavenumber component and $\hat{\bm{q}} = \hat{\bm{q}} (t, z)$ is a vector. Here, the normal mode of the disturbance is assumed as 
\begin{equation}
\bm{q}'(x,y,z,t) = \sum_{-\infty \leq k_x, k_y \leq \infty} \hat{\bm{q}}_k(z,t) \mathrm{e}^{i \bm{k} \cdot \bm{x}}
\end{equation}
where $\bm{k} = (k_x, k_y)$ is a wavenumber. Because we consider the quadratic nonlinear term, $\hat{N}_k$ can be represented as 
\begin{equation}
\hat{N}_k (t, \bm{q}') = \sum_{-\infty \leq l_x, l_y \leq \infty} \iint_{(x,y) \in [0,L_x] \times [0,L_y]} N (\hat{\bm{q}}_l \mathrm{e}^{i \bm{l} \cdot \bm{x}}, \hat{\bm{q}}_{k-l} \mathrm{e}^{i (\bm{k}-\bm{l}) \cdot \bm{x}}) \mathrm{e}^{- i \bm{k} \cdot \bm{x}} dx dy. 
\end{equation}
At the boundary of $z = \pm 1$, we assume a Dirichlet boundary condition detailed in the discussion below. We mainly consider the solution of $\hat{\bm{q}}_k = \hat{\bm{q}}_k (t, z)$ in (\ref{equ:non_dis_wave}). 

Considering a closed flow region, and assuming that the discretized linear operator is diagonalizable, any linear disturbance field can be represented as a linear combination of LST modes due to the completeness of the eigenmodes, as a conventional assumption in linear optimal perturbation analysis~\cite{hooper1996two}. Because of the non-normality of the linearized N--S operator in a finite-dimensional space (e.g., $\mathbb{R}^{M_{\mathrm{tot}}}$ with $N_t = K \times N_z$, where $K$ is the number of variables and $N_z$ is the number of discretization points in $z$ direction), the disturbance can be decomposed as
\begin{equation}
\hat{\bm{q}}_k (t, z) = \sum_{j=1}^{M_{\mathrm{tot}}} a_j^k (t) f_j^k (z), 
\end{equation}
where $f_j^k (z)$ is the $j$-th eigenvector corresponding to the eigenvalue $\sigma_j^k \in \mathbb{C}$ and $a_j^k (t) = P_j^k \hat{\bm{q}}_k (t)$ where $P_j^k = \langle f_j^{k*}, \cdot \rangle/ \langle f_j^{k*}, f_j^k \rangle$ is its amplitude. After taking projection and time-integration, the amplitude equations can be obtained as a Volterra integral equation as follows: 
\begin{align}
\label{eqn:volterra_int_equ}
a_j^k (t) = \eta_j^k (t) + \int_0^t \mathrm{e}^{(t-s) \sigma_j^k} P_j^k \hat{N}_k (s, \bm{q}') ds 
\end{align}
with $\eta_j^k (t) = \mathrm{e}^{t \sigma_j^k} a_j^k (0)$. The approximate solution of Eqn. (\ref{eqn:volterra_int_equ}) can be obtained in the Picard approximation when the nonlinear term is assumed to satisfy the local Lipschitz condition under the appropriate selection of disturbance growth to ensure the contraction mapping. In updating from the $n$-th step to $(n+1)$-th step in the Picard iteration, the equation becomes 
\begin{align}
a_j^{k, n+1} (t) &= \eta_j^k (t) + \int_0^t \mathrm{e}^{(t-s) \sigma_j^k} P_j^k \hat{N}_k (s, \bm{q}^{n}) ds
\end{align}
where $\bm{q}^{n}$ is reconstructed from $\{ a_j^{k, n} (t) \}$. 

\subsection{Physical observation for the assumptions}

The properties of the linear operator $\hat{L}_k$ have been studied in linear stability theory, and the negative real parts of the least stable eigenvalues depend especially on Reynolds number. By setting a sufficiently small Reynolds number, it is possible to assume that all the eigenmodes are linearly stable (i.e., $\operatorname{R} [\sigma^k_j] < 0$ for all $\bm{k}$) (e.g., for the incompressible limit of plane Poiseuille flow, $\mathrm{Re} < 5772$ satisfies this condition according to numerical simulations~\cite{orszag1971accurate}). In this case, when we define the linear amplification factor $G_k (t) := \max_{\hat{\bm{q}} (0)} || \hat{\bm{q}} (t) ||/ || \hat{\bm{q}} (0) ||$ under the linear time-evolution, the numerical simulation indicates that $G_k < C_k \mathrm{e}^{\alpha_k t} (\exists C_k > 0)$ where $\alpha_k = \max_j \operatorname{R} [\sigma^k_j] < 0$ for a linearly stable regime. Moreover, in the numerical simulation, it is found that the linear amplification attributed to the non-orthogonality of the modes is restricted to a finite number of modes~\cite{reddy1993energy, hanifi1996transient}. Physically, $G_k$ with a sufficiently large $k$ should decrease ($G_k \ll 1$ as $|\bm{k}| \to \infty$) due to the strong viscous dissipation. These properties indicate that the convergence of the Picard--Chebyshev method can be expected for sufficiently small disturbances to satisfy the local Lipschitz conditions in a subcritical regime, where the actual convergence will be discussed in Sec. \ref{sec:Numerical results} in comparison with DNS calculations. 

It should be noted that the present formulation fundamentally differs from the classical Stuart-Landau theory or multi-scale asymptotic expansions. While the standard weakly nonlinear theory relies on the existence of a spectral gap to assume the existence of center manifolds near the critical Reynolds number, it inherently fails to capture the linear amplification of disturbances originating from the strong non-normality of the system. The comparison of the present formulation and formal amplitude expansion of the disturbances will be discussed in a subsequent section. 

\subsection{Computational algorithm}

The computational algorithm is based on the Picard--Chebyshev method proposed by Fukushima~\cite{fukushima1997picard}. Let us assume that the initial disturbance field contains a finite number of discrete wavenumber components. With an application of the $n$-th Picard iteration, the entire set of modes in wavenumber space monotonically increases. The entire set in the $n$-th application of Picard iteration is denoted by $B_n$, and this sequence of sets constructs a monotonically increasing sequence determined from the Eqs. (\ref{equ:non_dis_wave}). For example, the first three sets in the case of a monochromatic wave are as follows: $B_1 = \{\bm{k}, -\bm{k} \}, B_2 = \{-2 \bm{k}, -\bm{k}, \bm{0}, \bm{k}, 2 \bm{k}\}, B_3 = \{ -4 \bm{k}, \dots, 4 \bm{k} \}$. 

For the evolution of disturbances $0 \leq t \leq t_f$ with a given external time $t_f > 0$, we introduce the transformation of variables $t = t_f (1 - \tau) / 2$ where $\tau = \cos \theta$ $(\theta \in [0, \pi])$. The projection of nonlinear terms to each mode is approximated by the $M_t$-th order Chebyshev polynomial expansions as 
\begin{equation}
P_j^k \hat{N}_k (s, \bm{q}^n) = \sum_{m=0}^{M_t} C_m^{k, j} T_m (\tau), 
\end{equation}
where $T_m (\tau) = \cos (m \theta)$ is the $m$-th order Chebyshev polynomial and $C_m^{k, j}$ are the coefficients. Then, $a_j^{k, n+1}-\eta_j^k$ can be expressed as 
\begin{equation}
a_j^{k, n+1} (t) -\eta_j^k (t) = \sum_{m=0}^{M_t} C_m^{k, j} I_m (t), \ I_m (t) = \int_0^t \mathrm{e}^{(t-s) \sigma_j^k} T_m (\tau(s)) ds
\end{equation}
where $I_m$ can be obtained as 
\begin{equation}
\frac{I_{m+1}}{m+1} - \frac{I_{m-1}}{m-1} = - \frac{4}{t_f \sigma_j^k} I_m + \frac{F_{m+1}}{m+1} - \frac{F_{m-1}}{m-1}, 
\end{equation}
with 
\begin{equation}
F_m = - \frac{1}{\sigma_j^k} (T_m (\tau(t)) - \mathrm{e}^{t \sigma_j^k} T_m (\tau(0))), 
\end{equation}
for $m>1$ using the identity $2 T_m (x) = T_{m+1}' (x)/(m+1) - T_{m-1}' (x)/(m-1)$. For $m=0$, $I_0 (t) = (\mathrm{e}^{t \sigma_j^k} - 1)/\sigma_j^k$ and 
\begin{equation}
I_1(t) = \frac{1}{\sigma_j^k} \left( \frac{2t}{t_f} - 1 + \mathrm{e}^{t \sigma_j^k} \right) - \frac{2}{t_f (\sigma_j^k)^2} (\mathrm{e}^{t \sigma_j^k} - 1). 
\end{equation}
These recurrence terms can be solved as a linear problem with a tridiagonal matrix. When $\sigma_j \approx 0$, this integral becomes numerically unstable, and it is replaced by an analytical integration based on the linear approximation of $\exp((t-s)\sigma_j^k) \approx 1.0 + (t-s)\sigma_j^k$. 

\section{Governing equations and numerical methods}
\label{sec:Governing equations and numerical methods}

\subsection{Governing equations and base flow field}

In this study, we consider the plane Poiseuille flow between plates at $z=\pm h$. The streamwise, spanwise, and wall-normal coordinates are denoted by $x, y$, and $z$, respectively. The flow variables are non-dimensionalized using the centerline density, the mean wall temperature of the two plates, channel half-width $h$, and a constant streamwise pressure gradient between the plates. The governing equations are the non-dimensional compressible N--S equations
\begin{align}
\partial_t \rho + u_j \partial_j \rho &= - \rho \partial_j u_j, \\ 
\rho \partial_t u_i + \rho u_j \partial_j u_i &= - C \partial_i (\rho T) + \partial_j \tau_{ij}, \\
\rho \partial_t T + \rho u_j \partial_j T &= - \tilde{\gamma} \rho T \partial_j u_j + A \partial_j (\kappa \partial_j T) + B \partial_j u_i \tau_{ij}, 
\end{align}
where $\rho, u_i, T, \gamma (= \tilde{\gamma}+1)$ are the density, velocity, temperature, and the specific heat ratio and $A = \gamma / (\mathrm{Pr} \mathrm{Re}), B=\gamma \tilde{\gamma} \mathrm{M}^2, C=1/(\gamma \mathrm{M}^2)$. The viscous stress tensor is given by
\begin{equation}
\tau_{ij} = D \mu (\partial_j u_i + \partial_i u_j - \frac{2}{3} \delta_{ij} \partial_k u_k ),
\end{equation}
with $D=1/\mathrm{Re}$ under Stokes' hypothesis for the bulk viscosity where $\mu, \kappa$ are the coefficient of viscosity and thermal conductivity. In this formulation, the non-dimensional parameters of Reynolds number $\mathrm{Re}$, Prandtl number $\mathrm{Pr}$, and Mach number $\mathrm{M}$ are defined based on the reference quantities. 

The reduction to one-dimensional equations is required to obtain the base flow solution $\bar{\bm{q}} = (\rho_0, U_0, 0, 0, T_0)^\top$. Under the parallel-flow assumption, these equations simplify to 
\begin{align}
\label{equ:base_ux}
2 + \mu_0 \frac{d^2 U_0}{d z^2} + \left( \frac{d \mu}{d T} \right)_0 \frac{d T_0}{d z} \frac{d U_0}{d z} &= 0,  \\
\label{equ:base_tem}
\frac{\mu_0}{\mathrm{Pr}} \frac{d^2 T_0}{d z^2} + \frac{1}{\mathrm{Pr}} \left( \frac{d \mu}{dT} \right)_0 \left( \frac{d T_0}{d z} \right)^2 + \mu_0 \tilde{\gamma} \mathrm{M}^2 \left( \frac{d U_0}{d z} \right)^2 &= 0. 
\end{align}
Assuming that the wall temperatures of the two plates are equal, boundary conditions at the walls are prescribed as 
\begin{equation}
\label{equ:bound_cond}
T_0 = 1, U_0 = 0,  
\end{equation}
at $z=\pm 1$. In the incompressible limit, this base flow simplifies to the well-known analytical solution for incompressible plane Poiseuille flow as $U_0 = 1 - z^2, T_0 = 1$. 

\subsection{Disturbance equations and weakly nonlinear analysis}
 
We introduce the state vector $\bm{q} (\bm{x}, t) = (\rho, u_x, u_y, u_z, T)^\top (\bm{x} \in \mathbb{R}^3, t \in \mathbb{R})$, and we assume the flow decomposition into steady base flow $\bar{\bm{q}} (z)$ and the perturbation $\bm{q}' (\bm{x}, t)$. The boundary condition of the disturbance field is $u_x' = u_y' = u_z' = T' = 0$ at $z= \pm 1$. 

Because the base flow is the steady solution of the compressible N-S equations, the governing equation can be symbolically represented as 
\begin{equation}
\label{equ:non_dis}
\partial_t \bm{q}' = L (\bar{\bm{q}}) \bm{q}' + N(\bm{q}', \bm{q}') + O(|| \bm{q}' ||^3)
\end{equation}
where $L$, $N$ denote the linearized N--S matrix and the first nonlinear term $N = N_1 (\bm{q}', \bm{q}')$, respectively. To avoid the discussion of infinite series expansion originated from $1/ (\rho + \rho')$ in the derivation of nonlinear disturbance field, we assume low Mach number flow to ignore the third and higher nonlinear terms in the followingdiscussion. Thus, this study treats the quadratic compressible N--S equations to eliminate higher order terms as (\ref{equ:equ_quad_dis_equ}). The detailed equations are summarized in Appendix \ref{sec:Nonlinear perturbation equations}. 

\subsection{Formulation as a weakly nonlinear theory (WNT)}
\label{sec:Formulation as a weakly stability theory}

In this subsection, we present a formulation of weakly nonlinear theory (WNT) tailored for direct comparison with the proposed P--C(n) method. While standard WNT models (e.g., Stuart--Landau equation) typically rely on the method of multiple scales, such formulation cannot be directly applied as a reference to evaluate the transient growth of finite-amplitude disturbances. 

To establish a rigorous reference based on amplitude expansion method, we derive a formulation to track a finite-amplitude monochromatic wave. Let $\epsilon \ll 1$ be a small amplitude parameter. Unlike in the previous formulation of the P--C method, we assume scale separation in the following formulation. We expand the disturbance around the fundamental wavenumber $\bm{k} = (k_x, k_y)$ as 
\begin{equation}
\label{equ:dis_sep}
q' = \epsilon [ \sum_{0 < j \leq M_{\mathrm{tot}}} (A_j \mathrm{e}^{\sigma_j t} + \epsilon^2 a_j (t)) \psi_{11}^j + \epsilon^2 \psi_{13} ] \mathrm{e}^{i \bm{k} \cdot \bm{x}} + \epsilon^2 (\psi_{22} \mathrm{e}^{2 i \bm{k} \cdot \bm{x}} + \psi_{02}) + \mathrm{c.c.}
\end{equation}
where the superscript $k$ is omitted and $\mathrm{c.c.}$ denotes the complex conjugates. Here, $M_{\mathrm{tot}}$ represents the total number of modes. The spatial shapes $\psi_{11}, \psi_{13}, \psi_{22}, \psi_{02}$ correspond to the fundamental wave, its third-order spatial correction, the second-harmonic, and the mean-flow distortion, respectively. 

To guarantee that the third-order spatial correction $\psi_{13}$ remains bounded (i.e., to eliminate the secular term arising from resonance with the fundamental linear operator $\hat{L}_k$), we impose the solvability condition using the adjoint eigenmodes $(\psi_{11}^j)^{\dagger}$: 
\begin{equation}
\langle (\psi_{11}^j)^{\dagger}, \psi_{13} \rangle = 0 \quad \text{for all} \quad 0 < j \leq M_{\mathrm{tot}}. 
\end{equation}

By equating terms of corresponding powers of the amplitude, the system decouples into governing equations for spatial modes and amplitude corrections: 
\begin{align}
\label{equ:wn_model_1}
\partial_t \psi_{02} &= \hat{L}_0 (\bar{q}) \psi_{02} + 2 N_1 (\psi_{11}, \psi_{-11}), \\
\label{equ:wn_model_2}
\partial_t \psi_{22} &= \hat{L}_{2k} (\bar{q}) \psi_{22} + N_1 (\psi_{11}, \psi_{11}), \\
\label{equ:wn_model_3}
\partial_t \psi_{13} &= \hat{L}_k (\bar{q}) \psi_{13} + R',  \\
\label{equ:solvability}
 \partial_t a_j (t) &= \sigma_j a_j (t) + \hat{R}_j
\end{align}
where $\psi_{-11} = (\psi_{11})^*$. Here, $R$ represents the total nonlinear forcing generated by fundamental and harmonic interactions: 
\begin{equation}
R = 2 N_1 (\psi_{11}, \psi_{02}) + 2 N_1 (\psi_{-11}, \psi_{22}) + 3 N_2 (\psi_{11}, \psi_{11}, \psi_{-11})
\end{equation}
with $\hat{R}_j = \langle (\psi_{11}^j)^{\dagger}, R \rangle/ \langle (\psi_{11}^j)^{\dagger}, \psi_{11}^j \rangle$ and $R' = R - \sum_{0 < j \leq M_{\mathrm{tot}}} \hat{R}_j \psi_{11}^j$. Note that the third-order nonlinear terms in this formulation are retained for comparison with the P--C(n) method based on quadratic equations. 

\subsection{Numerical discretization method and flow settings}

\subsubsection{Numerical method for P--C and WNT analysis}

The flow field is configured for the subcritical condition of plane Poiseuille flow at $\mathrm{M}=0.3, \mathrm{Re}=3000, \mathrm{Pr}=0.72$. This condition matches the flow setting used by \cite{reddy1993energy, hooper1996two} for linear non-modal studies. For these flow settings, we employed a sixth-order compact difference method~\cite{lele1992compact} to approximate the coefficients of differential operators. The wall-normal domain was discretized using $256$ points and clustered near the wall with a minimum grid spacing $\Delta z = 3.99 \times 10^{-3}$. The generalized eigenvalue problem was solved to obtain the eigenpairs using the Schur complement to satisfy boundary conditions. In actual calculations of WNT and Picard--Chebyshev method, all $1272$ modes were used at each wavenumber component in $B_n$ (corresponding to the interior degrees of freedom) to ensure full-rankness. Adjoint modes were directly obtained from the generalized eigenvalue problem, after checking that the bi-orthogonality between eigenmodes and adjoint modes satisfied the bi-orthogonality condition within machine precision. 

\begin{figure}
  \centering
  \begin{minipage}{0.495\textwidth}
    \centering
    \includegraphics[width=6cm]{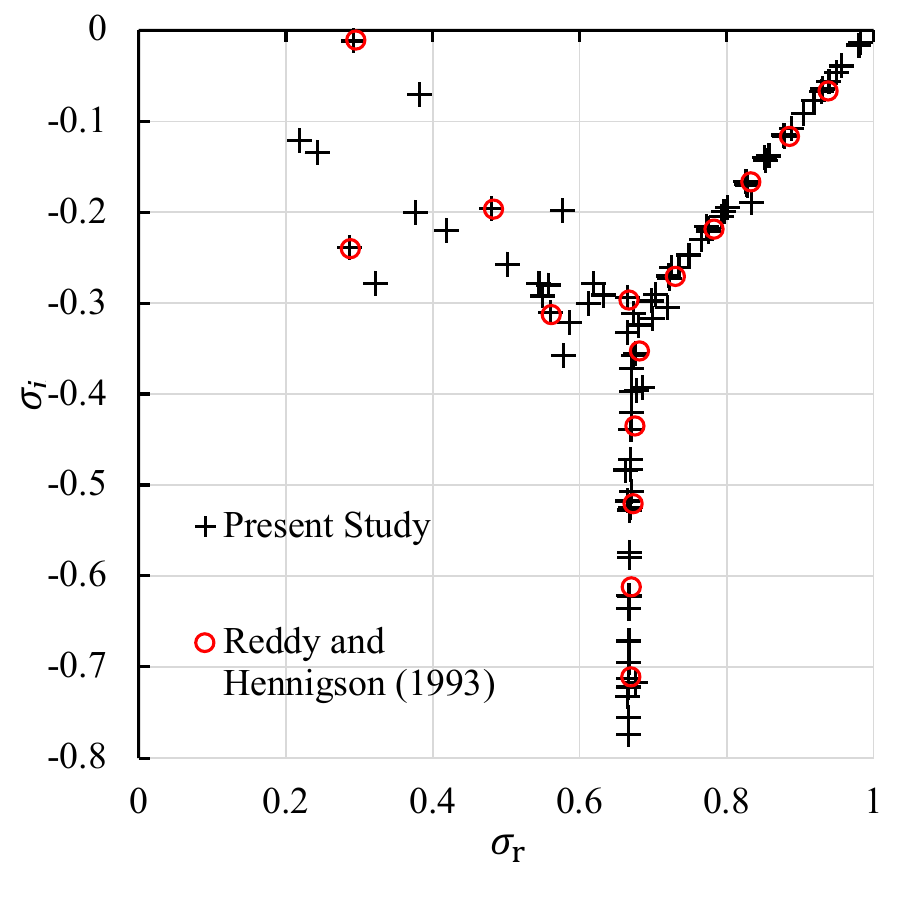}
    \subcaption{Spectrum distribution}
  \end{minipage} 
  \begin{minipage}{0.495\textwidth}
    \centering
    \includegraphics[width=6cm]{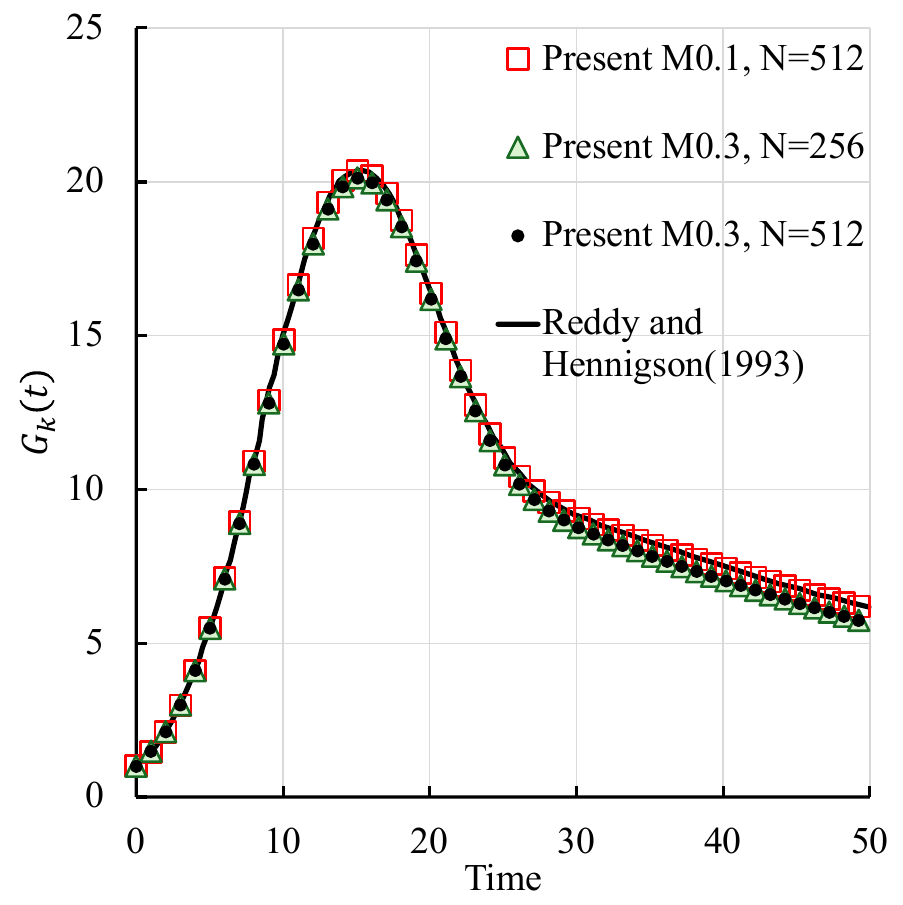}
    \subcaption{Linear amplification rate}
  \end{minipage}
  \caption{
(a) Eigenvalue distribution and (b) linear amplification rate at $\mathrm{M}=0.1, 0.3$ and $\mathrm{Re}=3000, k_x = 1, k_y = 0$ in comparison with literature values~\cite{reddy1993energy}. 
  }
  \label{Fig:eigenvalues}
\end{figure}

The numerical validation is presented in Fig. \ref{Fig:eigenvalues} against the previous study of linear stability analysis and linear optimal perturbation analysis~\cite{reddy1993energy} for incompressible plane-Poiseuille flow. The present study accurately reproduces the known eigenvalues from the previous study, along with additional values arising from the extension to a 3D compressible model. The linear amplification rate, calculated following the standard formulation~\cite{reddy1993energy, hooper1996two}, shows excellent agreement with the literature values regardless of differences in the numerical discretization approach. The same figure also shows the grid convergence of linear amplification, indicating the sufficient resolution to capture the amplification. For large evaluation times exceeding $30$, the calculation results at $\mathrm{M}=0.3$ show a slight underestimation due to compressible effects, whereas the results at $\mathrm{M}=0.1$ confirm convergence in the low-Mach-number limit, where the fine mesh ($N_z=512$) is employed to capture the acoustic modes in low Mach numbers. 

\subsubsection{DNS setup for reference comparison}

For ground-truth data, we also conduct DNS simulations to track the temporal evolution of the disturbances in Secs. \ref{sec:Properties of Picard-Chebyshev method in monochromatic wave} and \ref{sec:Weakly nonlinear optimal perturbations}, where the former section considers monochromatic waves and the latter considering polychromatic waves. In the former case, we consider the computational domain of $[0,4\pi]$ for the streamwise and $[0, 2\pi]$ for the spanwise direction, which spans two wavelengths of the fundamental mode, $k_x = 1$, as a main target of the present calculation. The numerical grid points for $x, y$ are both $256$ and the wall-normal grid points distribution is common to the WNLOP analysis, where the total grid number is about $17$ million points. In the latter case, we extend the computational domain into $[0, 8 \pi]$ for streamwise and $[0, 4 \pi]$ for spanwise direction to consider the two fundamental waves of $\bm{k} = (0, 1)$ and $(0.5, 0)$. With an extension of computational domain, the grid points also increase into $512$ for both $x, y$ directions to keep the same resolution for all the calculations, where the total number of grid points is $67$ million points. 

The numerical discretization method is common to the P--C and WNT analyses and time-stepping procedure is performed using a third-order Total Variation Diminishing (TVD) Runge-Kutta method~\cite{gottlieb1998total}. In DNS, we applied the $10$th-order compact filtering method~\cite{gaitonde2000pade} with $\alpha_f=0.45$. This filtering procedure was also utilized during the evaluation of nonlinear terms in WNLOP to suppress the aliasing error.

\section{Numerical results}
\label{sec:Numerical results}

\subsection{Properties of Picard-Chebyshev method for a monochromatic wave}
\label{sec:Properties of Picard-Chebyshev method in monochromatic wave}

\begin{figure}
\centering
\includegraphics[width=8cm]{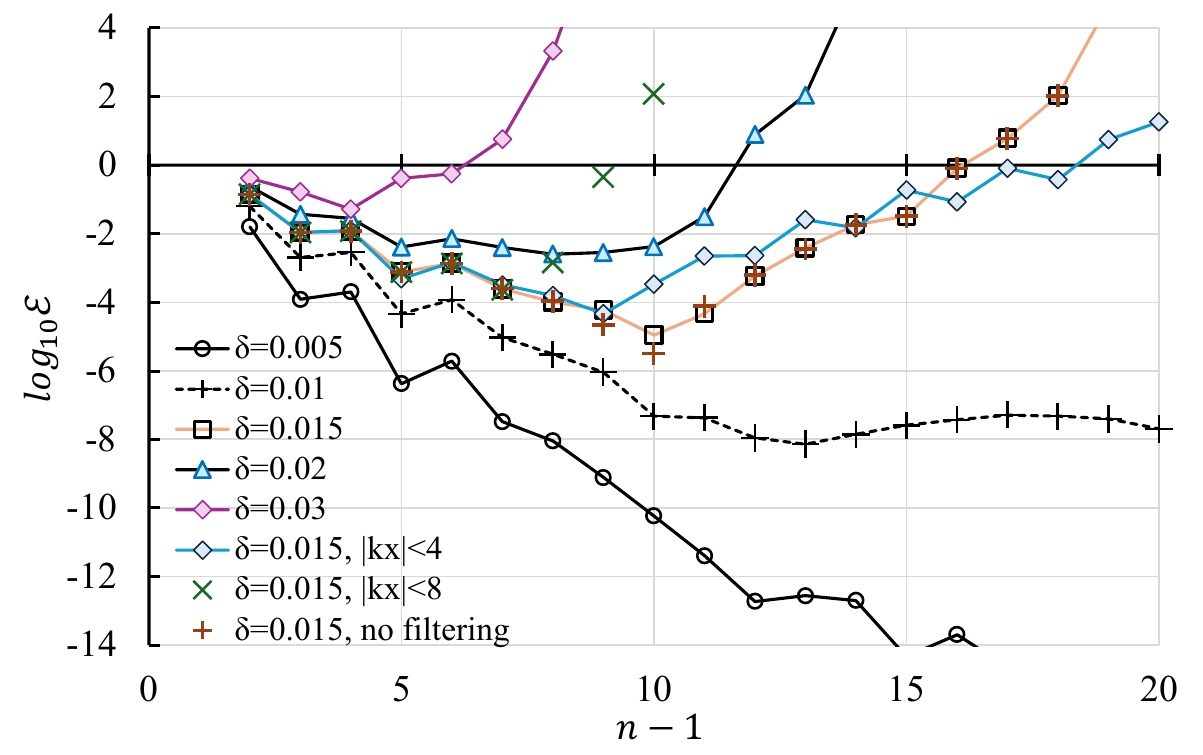}
\caption{
Relative residual $\mathcal{E}$ in each iteration of the P--C method, initiated from the linear optimal perturbation at $k_x = 1, k_y = 0, t_f = 15.0$. Unless otherwise specified in the legend, the maximum wavenumber is restricted to $|k_x| < 6$, and compact filtering is used to eliminate aliasing errors. 
}
\label{Fig:pic_conv_PC}
\end{figure}

We first evaluate the convergence of the P--C method using the relative residual defined as 
\begin{equation}
\mathcal{E} (n) = \frac{|| \bm{a}^{n, i} - \bm{a}^{n+1, i} ||_{l^2}}{|| \bm{a}^{n, i} ||_{l^2}}
\end{equation}
for the wave $\bm{k}_i$ contained in $B_n$. Figure \ref{Fig:pic_conv_PC} shows the relative residual during P--C(n) iterations initiated from the linear optimal perturbation with a finite initial disturbance amplitude $\delta_0$, as measured by the Mack norm~\cite{hanifi1996transient}. The calculation was conducted for a monochromatic wave $(k_x, k_y) = (1, 0)$ and $t_f = 15$, and the temporal discretization in the P--C method is $M_t=512$ points after confirming no significant difference with $M_t=1024$ points. For a small initial disturbance amplitude, the method exhibits robust convergence, where increasing $n$ consistently reduces $\mathcal{E}$. This confirms the fundamental validity of convergence in the P--C scheme under linear or weakly nonlinear conditions. However, as the initial disturbance amplitude increases, the iteration begins to diverge regardless of the early decrease of relative residuals in P--C iterations due to the combined limitations of the contraction mapping and numerical simulation. 

To clarify the cause of this numerical instability, we examine the effect of wavenumber truncation at a fixed amplitude $(\delta_0 = 0.015)$. Increasing the truncation wavenumber leads to numerical divergence, which can be attributed to the ill-conditioning of the matrix at high wavenumbers, where nearly parallel spurious modes are formed and amplify numerical noise even when the linear amplification is small due to viscous dissipation in high wavenumber components (e.g., the cases of $|k_x|<4$ and $|k_x| < 8$ shown in Fig. \ref{Fig:pic_conv_PC}). While truncating these high-wavenumber components restores convergence, this suppression remains effective only until the unphysical accumulation of wave energy eventually destabilizes the numerical solution. From a comparative numerical experiment on the application of compact filtering, the effect of aliasing errors originating from nonlinear terms is negligible, suggesting that the interruption of energy transfer to higher modes is the primary trigger of this instability. 

\begin{figure}
\centering
\includegraphics[width=8cm]{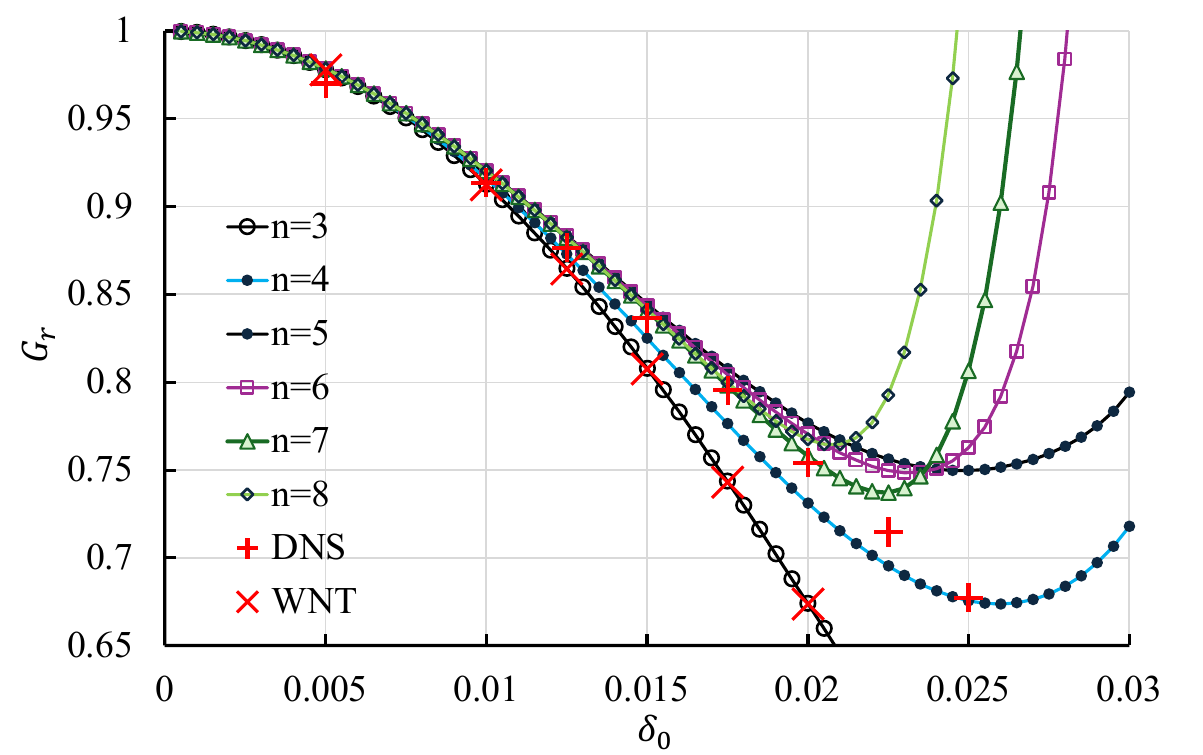}
\caption{
Comparison of disturbance amplification predictions as a function of the initial disturbance amplitude, normalized by the linear growth rate and initiated from the LOP condition estimated at $t_f = 15$. On the vertical axis, $G_r$ is defined as $G_r = || \hat{\bm{q}} (t_f) ||_e/(\delta_0 G_k (t_f))$. 
}
\label{Fig:pic_comp_PC}
\end{figure}

Figure \ref{Fig:pic_comp_PC} compares the prediction accuracy between WNT and the P--C models with respect to nonlinear suppression of linear optimal perturbation at finite amplitude against the DNS calculation. According to the DNS calculation, the growth rate is reduced by $10 \%$ at $\delta_0 = 0.01$ and by $25 \%$ at $0.02$ compared to the LOP prediction. Both models appropriately predict the suppression of linear amplification at finite disturbance amplitude; however, an improvement of predictive accuracy is recognized in the P--C method for $\delta_0 > 0.01$, where the WNT model overestimates the damping of disturbance growth. In contrast, the P--C method maintains a high predictive accuracy against DNS calculations until the aforementioned high-wavenumber components finally destabilize the calculations. 

\begin{figure}
  \centering
  \begin{minipage}{0.495\textwidth}
    \centering
    \includegraphics[width=6cm]{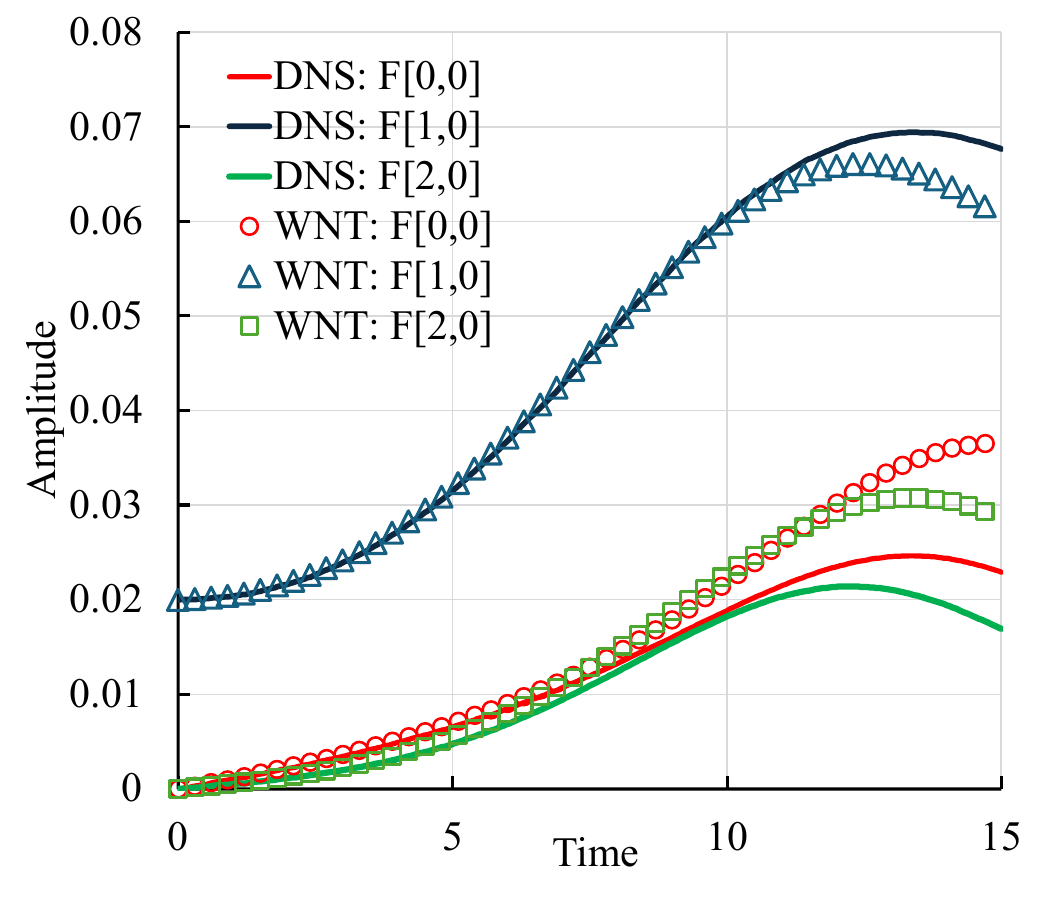}
    \subcaption{WNT/DNS comparison}
  \end{minipage} 
  \begin{minipage}{0.495\textwidth}
    \centering
    \includegraphics[width=6cm]{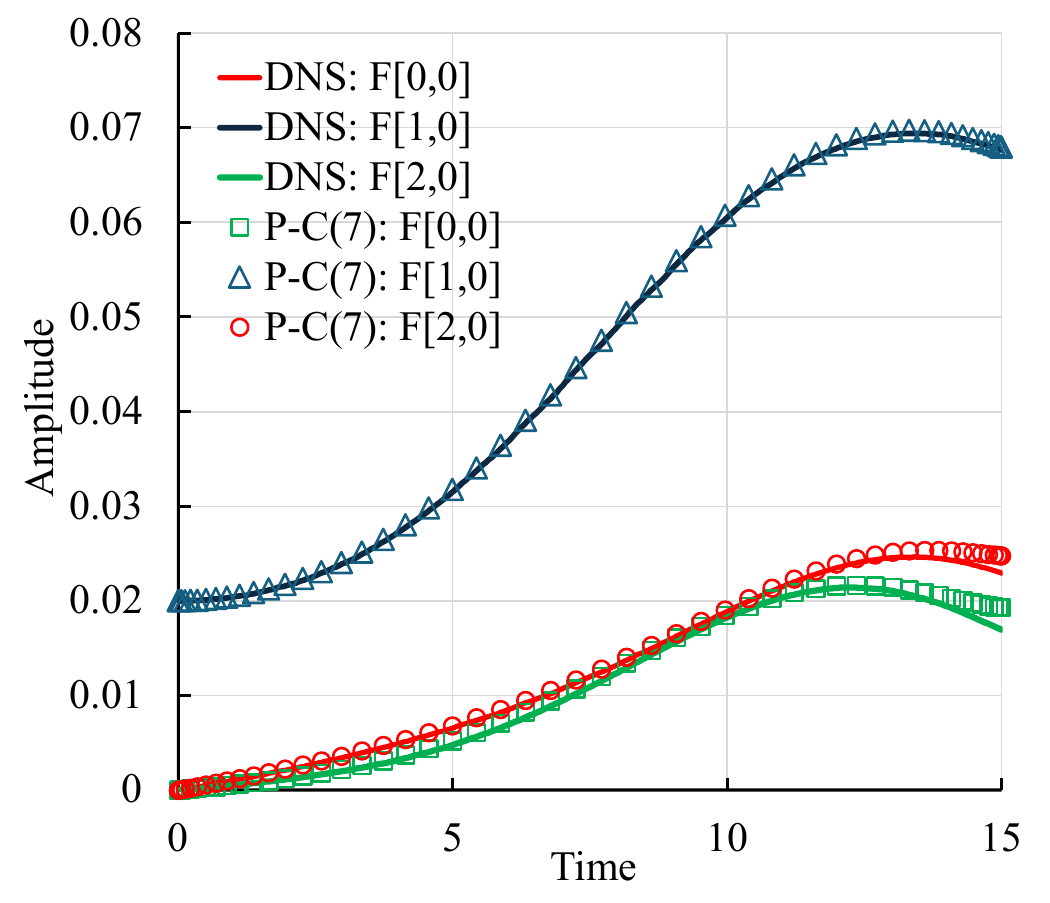}
    \subcaption{P--C(7)/DNS comparison}
  \end{minipage}
  \caption{
Comparison of WNT and P--C (7) methods against DNS calculations at $k_x = 1, k_y = 0, t_f = 15, \delta_0 = 0.02$. In each figure, $F[k_x, k_y]$ represents the amplitude of each wavenumber component. 
  }
  \label{Fig:pic_comp_PC_DNS}
\end{figure}

The numerical results support the theoretical consistency of the WNT model with the third-order truncated model in the P--C method, demonstrating a good agreement between the two results despite their different formulations and time-stepping approaches. In the third-order truncated model, the P--C method incorporates the correlation of fundamental waves from the mean flow distortion and double harmonic waves from the second-order approximation, which is identical to that in the WNT formulation. The higher-order approximation of the P--C model correlates the evolution of the mean flow distortion and double harmonic waves, demonstrating asymptotic convergence to the DNS results. Figure \ref{Fig:pic_comp_PC_DNS} shows the time evolution of each wavenumber component in different models. The higher-order correlation between mean flow distortion and second harmonic components suppresses the overestimation of their evolution, thereby improving predictive accuracy. However, as discussed in Fig. \ref{Fig:pic_comp_PC}, the stronger nonlinear regime causes the overestimation of higher harmonic wave components and eventual calculation divergence. 

\subsection{Weakly nonlinear optimal perturbations}
\label{sec:Weakly nonlinear optimal perturbations}

The primary application of the formulation is to investigate the nonlinear interaction of disturbances in the early stages of turbulent transition. We consider the problem of obtaining an optimal solution to trigger secondary disturbances by maximizing the amplitude of disturbances at evaluation time $t_f$. When $B_1$ contains two discrete wavenumber components, namely $\hat{\bm{q}}^1, \hat{\bm{q}}^2$ at wavenumber $\bm{k}_1, \bm{k}_2$, and their complex conjugates, this problem can be defined to maximize
\begin{equation}
\mathcal{L} = \sum_{\hat{\bm{q}}^N \in B_N} || \hat{\bm{q}}^N (t_f) ||_e^2 
\end{equation}
after $N$-steps of Picard iterations under the constraint conditions of $|| \hat{\bm{q}}^1 (t_0) ||_e = \delta_1$ and $|| \hat{\bm{q}}^2 (t_0) ||_e = \delta_2$. Here, $\delta_1, \delta_2 > 0$ are the given initial disturbance amplitudes of each wavenumber component. Since the initial condition is identical in Picard iterations, it is sufficient to set the constraint condition for the total amplitude of $\{ a_j (0) \}_{j=0,1,\dots, M}$ in $B_1$. 

We assume that $\hat{\bm{q}}^1$ is fixed as a solution of streamwise linear optimal perturbation representing the streak structure, and we consider an optimization problem for $\hat{\bm{q}}^2$, which reformulates the problem as finding numerical solutions leading to turbulent transition initiated from a streak. According to the linear optimal perturbation analysis, the optimal evaluation time and its linear amplification at $(k_x, k_y) = (0, 1)$ are calculated as ${t_f}_{\mathrm{opt}} \approx 270$ at $\mathrm{M}=0.3, \mathrm{Re}=3000$, and $t_f = 250$ is chosen with $G (t_f = 250) = 928.86$ for the primary disturbance of $\hat{\bm{q}}^1$. The evolution of the primary wavenumber component $\hat{\bm{q}}^1$ is initiated at $t=-50$ and the second wavenumber component $\hat{\bm{q}}^2$ is added at $t=0$. This procedure aims to avoid applying secondary disturbances to the early formation stage of linear optimal perturbation into a streak structure, because the initial stage of LOP formulation is governed by the phase adjustment of eigenvectors rather than the physical evolution of streak structure. In the optimization calculation, the discrete adjoint method is employed to determine the search direction, after carefully investigating the numerical accuracy of the discrete adjoint vector. For the stepping determination, the momentum-based approach for step width in Pringle et al.~\cite{pringle2012minimal} is employed. 

After the initial oscillation in optimization starting from random noise, the solution rapidly converges into a maximum solution after tens of steps. Except for the numerical divergence of the model, the scheme successfully produces the converged solution. The convergence criterion is set to $||\nabla \mathcal{L}|| < 10^{-5}$ based on the effect on the evaluation function. All the calculations are conducted at $\bm{k}_1 = (0, 1), \bm{k}_2 = (0.5, 0)$, and $N=5$. The optimization conditions are set to $\delta_2 < 0.01$ and $t_f = 50, 75, 100, 150$. 

\subsubsection{Identified initial perturbation structure}

\begin{figure}
\centering
\includegraphics[width=7cm]{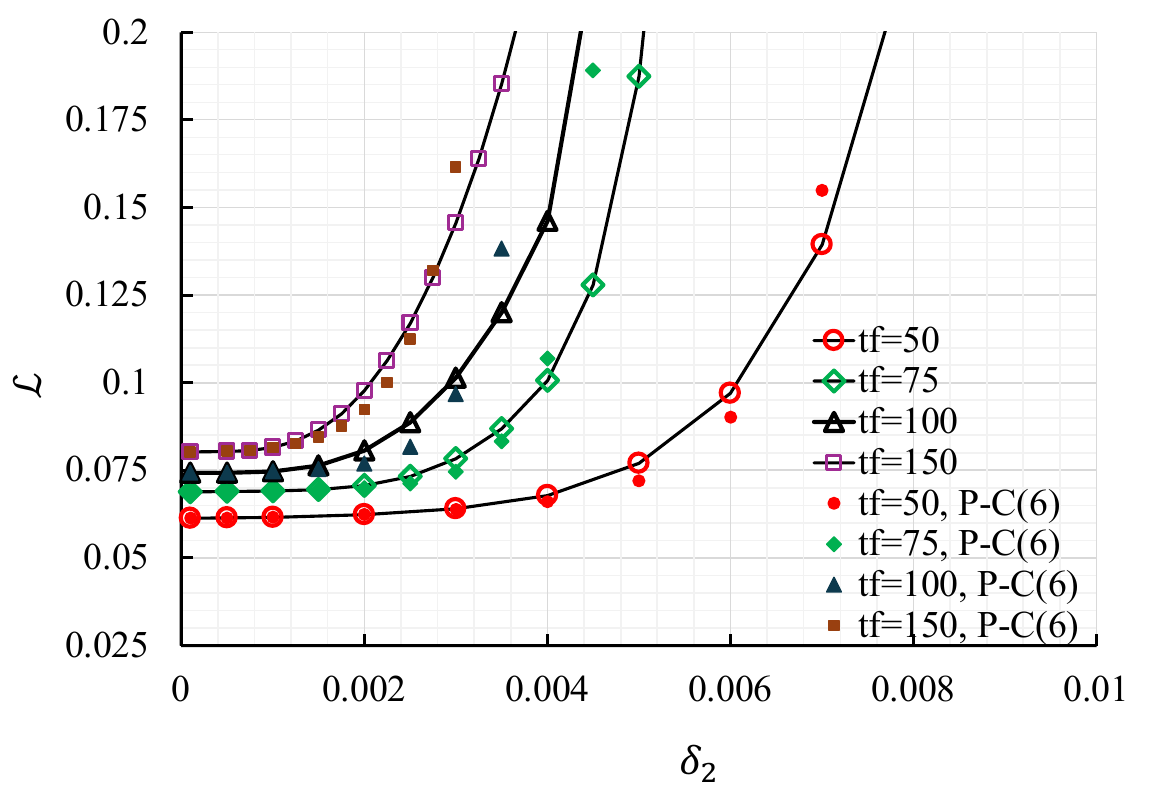}
\caption{
Optimized evaluation function $\mathcal{L}$ against $\delta_2$ and $t_f$ using the P--C(5) calculation at $\delta_1 = 0.002$. The calculated results with the P--C(6) are presented with filled symbols. 
}
\label{Fig:pic_delta2_L}
\end{figure}

Figure \ref{Fig:pic_delta2_L} shows the optimized value of $\mathcal{L}$ as a function of $t_f$ and $\delta_2$. For small $\delta_2$ at each $t_f$ condition, $\mathcal{L}$ approaches a constant value, which depends only on the amplification and the distortion of the streak in the first mode. The nonlinear interactions among the streak, the planar wave, and the subsequent excited waves generate a significant increase in $\mathcal{L}$ with increasing $\delta_2$ and $t_f$, where the appropriately selected WNLOP induces the laminar-to-turbulent transition between $t=50$ and $t=150$, as shown in the subsequent detailed results. In the same figure, the prediction of the evaluation function at P--C(5) compared to that at P--C(6) shows only negligible differences for small $\delta_2$ and begins to overestimate with increasing $\delta_2$, and exhibits a non-convergence behavior for $\mathcal{L} > 0.1$. This behavior is common to Fig. \ref{Fig:pic_comp_PC} with respect to the predictive limitations for monochromatic waves. Thus, the accurate prediction of WNLOP with a consideration of nonlinear interactions of polychromatic waves is restricted to a weakly nonlinear regime before the highly nonlinear regime leads to the spurious amplification of high wavenumber components. 

\begin{figure}
\centering
\includegraphics[width=12cm]{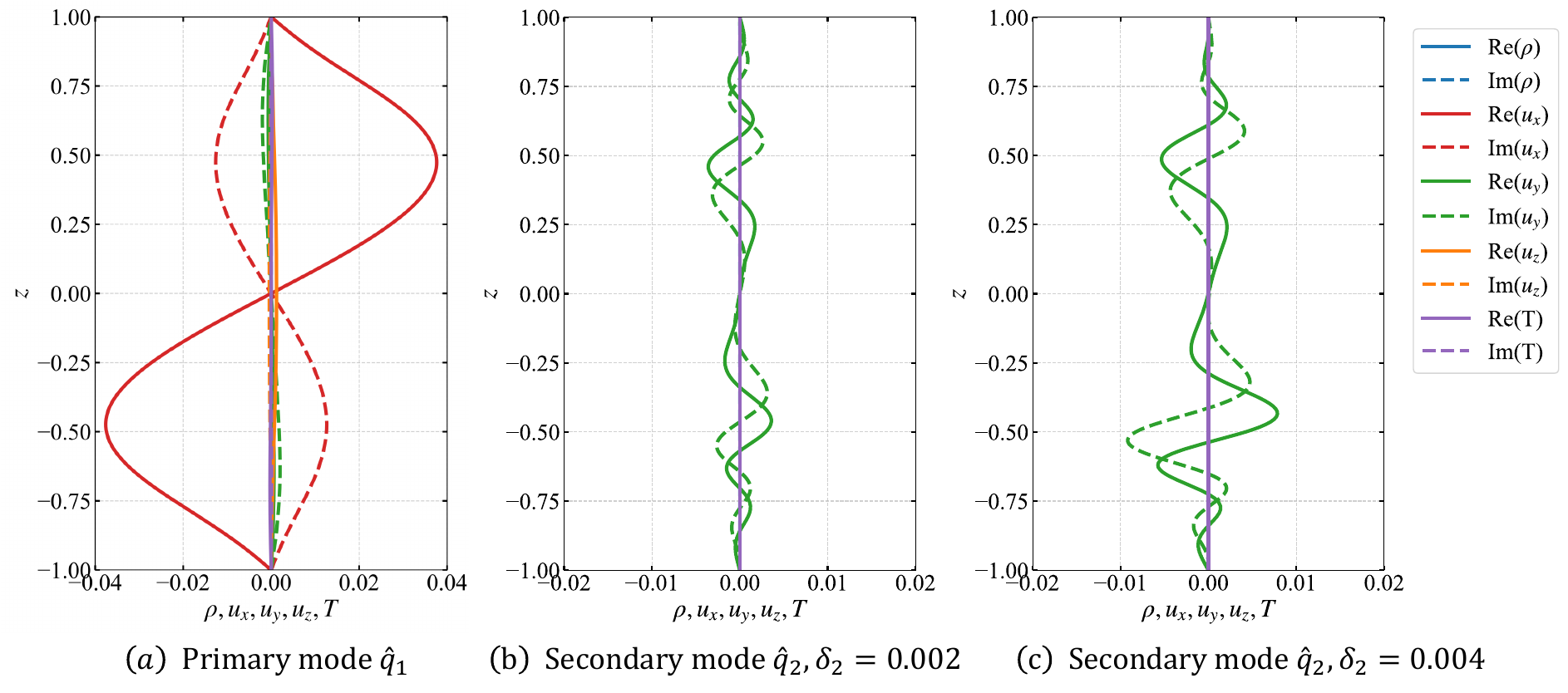}
\caption{
Wall-normal distribution of (a) the primary mode $\hat{\bm{q}}^1$ and (b, c) the secondary mode $\hat{\bm{q}}^2$ at $t=50$. Note that the horizontal axes have different scales between the primary and secondary modes. 
}
\label{Fig:profiles_combined}
\end{figure}

Figure \ref{Fig:profiles_combined} shows the wall-normal distribution of the primary and secondary modes. For a short evaluation time and a small $\delta_2$, the profile of $\hat{\bm{q}}^2$ shows a symmetric distribution with respect to $z=0$, which is the same as that of the primary mode. With an increase in $\delta_2$, the identified $\hat{\bm{q}}^2$ shows an asymmetric distribution with a local concentration of amplitude in the wall-normal direction. The maximum concentration is around $z=-0.4$ in both cases and the concentration near upper or lower boundary depends on the initial guess of the optimization calculation due to the point symmetry of base flow and $\hat{\bm{q}}^1$. The exact location of the maximum amplitude depends on the optimization conditions. The secondary mode is mainly composed of spanwise velocity fluctuations, while the other components have negligible amplitudes, comparable to machine error. Because $\hat{\bm{q}}^2$ is a function of $z$ with wavenumber components $k_x = 0.5, k_y=0$, this distribution generates streamwise and wall-normal vorticities ($\omega_x, \omega_z$) within the flow field. The phase difference between the real and imaginary components in $u_y$ shows a $\pi$ phase-shift between the top and bottom boundary walls, indicating a planar wave representing the concentration and diffusion at the center between two parallel plates. These spatial distributions were commonly observed in the calculated optimal solutions of $\hat{\bm{q}}^2$. 

\subsubsection{Transition process}

\begin{figure}
  \centering
  \begin{minipage}{0.495\textwidth}
    \centering
    \includegraphics[width=6cm]{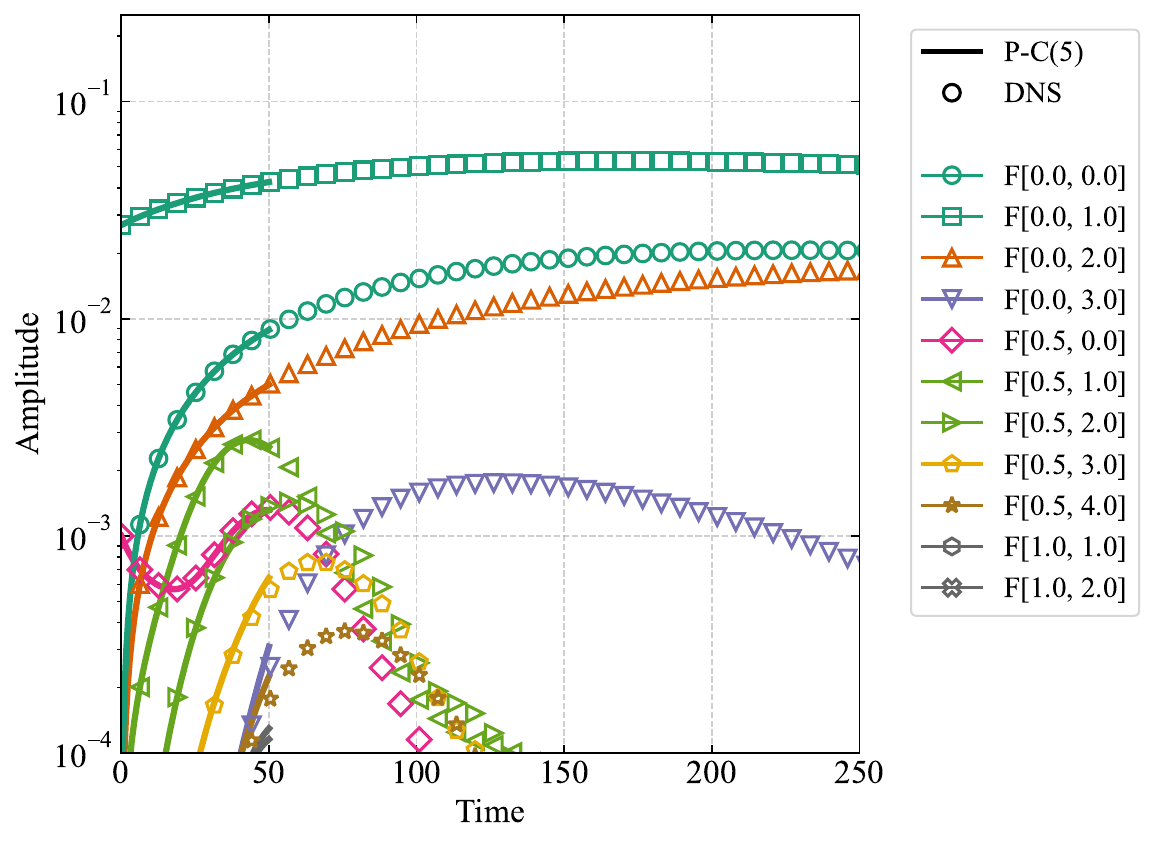}
    \subcaption{$\delta_2 = 0.001$}
  \end{minipage} 
  \begin{minipage}{0.495\textwidth}
    \centering
    \includegraphics[width=6cm]{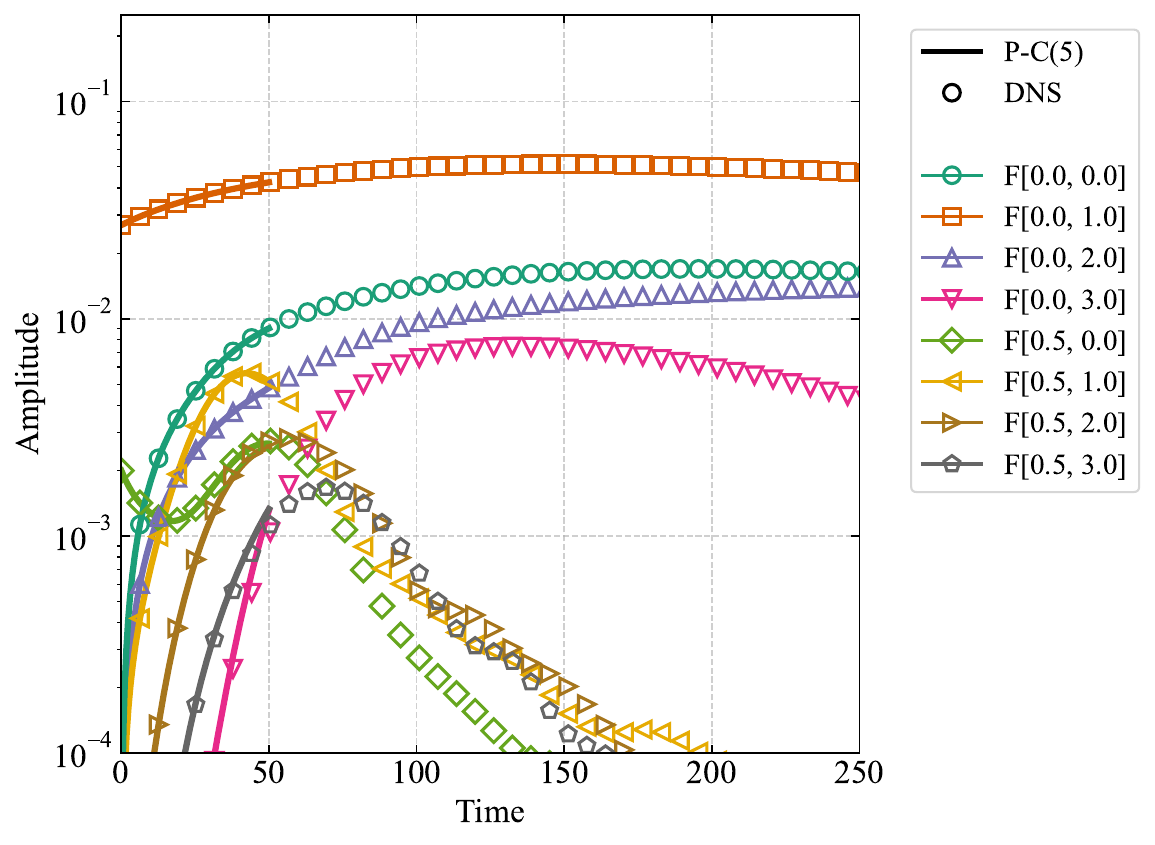}
    \subcaption{$\delta_2 = 0.002$}
  \end{minipage} \\
  \begin{minipage}{0.495\textwidth}
    \centering
    \includegraphics[width=6cm]{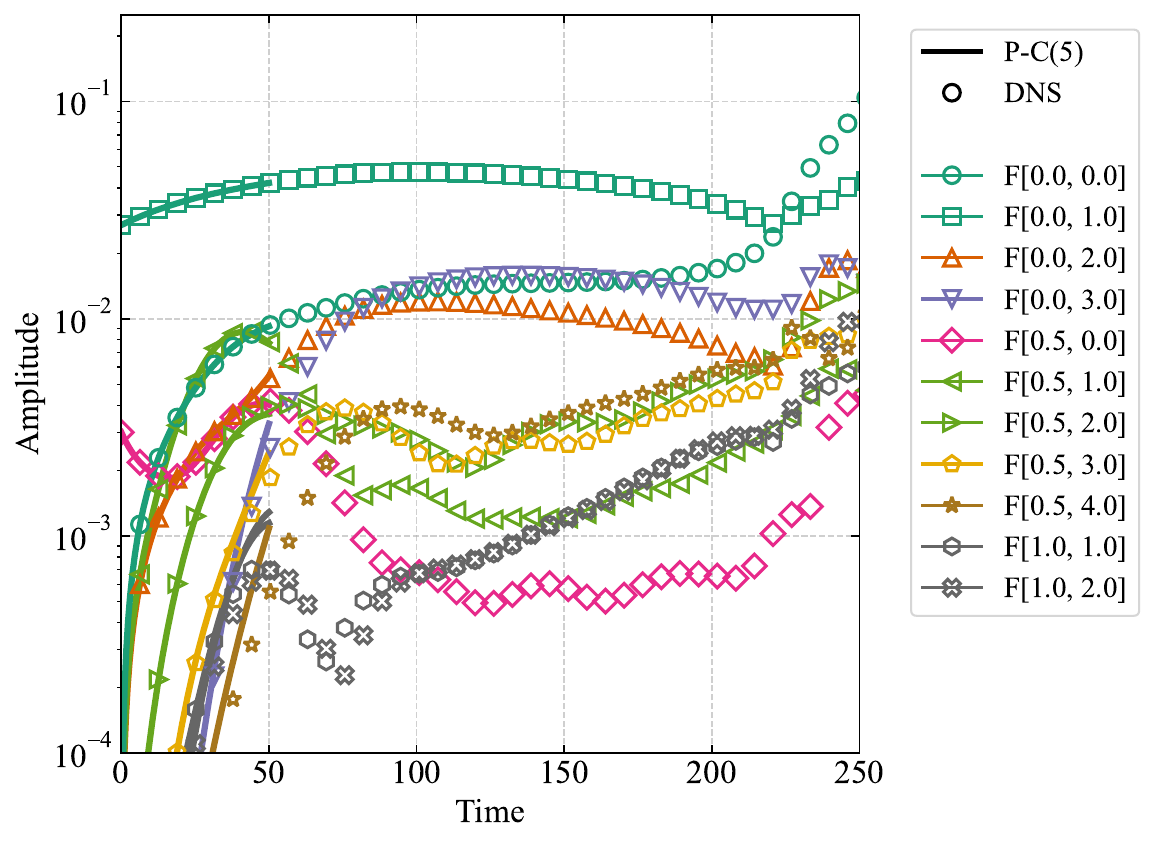}
    \subcaption{$\delta_2 = 0.003$}
  \end{minipage}  
  \begin{minipage}{0.495\textwidth}
    \centering
    \includegraphics[width=6cm]{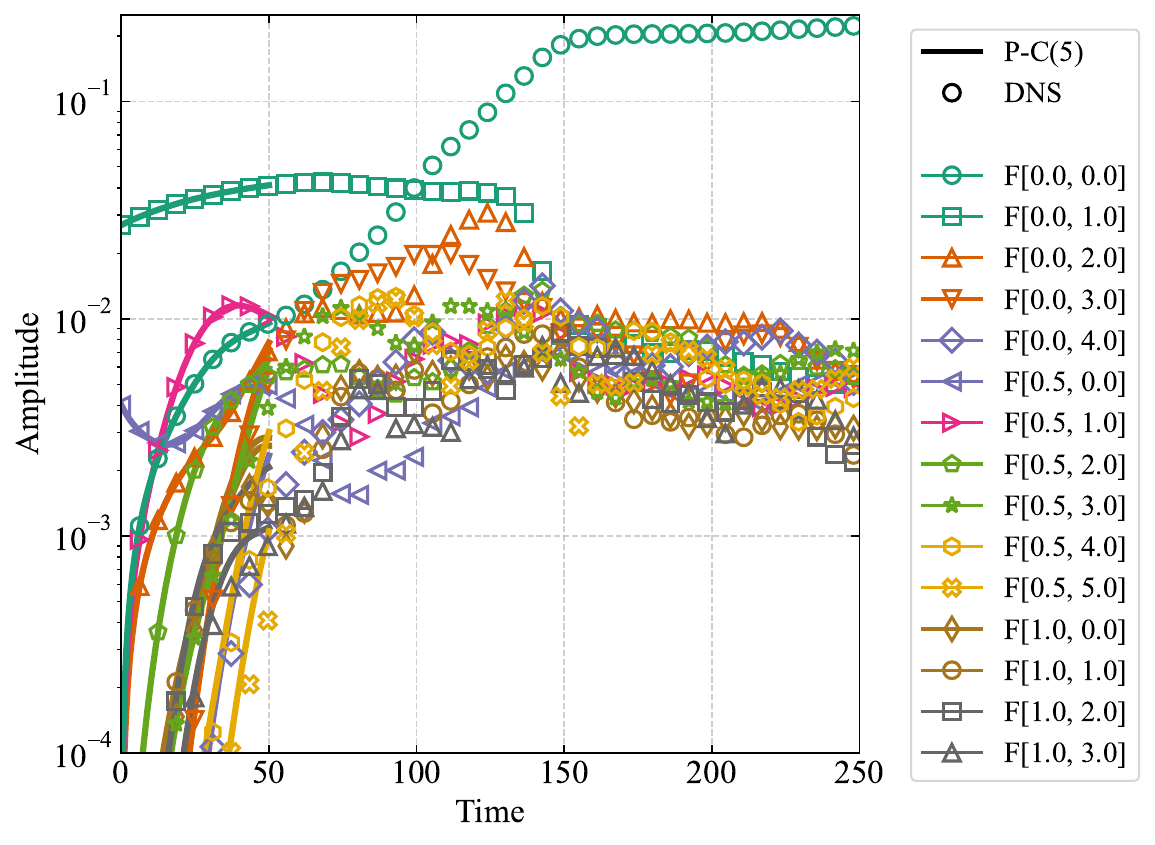}
    \subcaption{$\delta_2 = 0.004$}
  \end{minipage}
  \caption{
Comparison of the P--C(5) and DNS calculations for the optimized WNLOP at $t_f = 50$. The solid lines represent the P--C(5) calculations and symbols represent DNS calculations for the same wavenumber components using the same colors, where the threshold amplitudes for visualization are $F[k_x, k_y] > 10^{-3}$ ($10^{-4}$ at $\delta_2 = 0.001$), with only components for $k_x, k_y>0$ presented. 
  }
  \label{Fig:pic_freq_d004_T0050_N0200_M0200_kx050}
\end{figure}

Figure \ref{Fig:pic_freq_d004_T0050_N0200_M0200_kx050} shows the time-evolution of each wavenumber component initiating from the identified WNLOP, predicted in the P--C(5) and DNS calculations. The results of the P--C(5) calculation are plotted up to the evaluation time $t=50$, where the RMS errors between the P--C(5) and DNS are $4.53 \times 10^{-5}, 1.66 \times 10^{-4}, 4.47 \times 10^{-4}, 1.23 \times 10^{-3}$ at $\delta_2 = 0.001, 0.002, 0.003, 0.004$, respectively. These small RMS error values indicate that the P--C(5) calculation in WNLOP optimization accurately predicts the early time-evolution of the disturbances, although the predictive accuracy at $\delta_2 = 0.004$ can be improved with further P--C iterations (e.g., $5.4 \times 10^{-4}$ with P--C(7) at $\delta_2 = 0.004$). These results demonstrate that the predictive accuracy of the WNLOP analysis for polychromatic waves is sustained in comparison with DNS calculations when the P--C iteration converges well, as presented in Fig. \ref{Fig:pic_delta2_L}. 

In early time evolution, the nonlinear interactions between the streak and the planar wave initially generate the oblique wave at $\bm{k} = (0.5, 1)$, along with wave components at $\bm{k} = (0, 0), (0, 2), (0, 3)$ originating mainly from the self-interaction of the streak. Although this excitation of the oblique wave is commonly observed in Figs. \ref{Fig:pic_freq_d004_T0050_N0200_M0200_kx050} (a) - (d), the amplitude of the excited oblique wave determines the characteristics of the subsequent time evolution of the disturbances leading to the turbulence transition. A comparison among (a) - (d) shows that the excited amplitude of the oblique wave dictates the amplitude of waves in further nonlinear interactions, which include a group of wavenumbers at $\bm{k} = (0.5, 1), (0.5, 2), (0.5, 3)$ and subsequent modes at $\bm{k}=(1, 1), (1, 2)$, as presented in Fig. \ref{Fig:pic_freq_d004_T0050_N0200_M0200_kx050} (c). When the amplification of this group of wavenumbers decays during time evolution, the flow field contains only the harmonic waves from the streak, indicating that the numerical calculation tracks a simple distortion of the streak due to its finite amplitude. When the excited oblique wave is strong enough to generate further nonlinear interactions, the flow eventually reaches a stable turbulent state characterized by stable mean flow distortion amplitude at $\bm{k} = (0, 0)$ and the excitation of a large number of wavenumber components (see Figs. \ref{Fig:pic_freq_d004_T0050_N0200_M0200_kx050} (c), (d)). This evolution of wavenumber components is mainly attributed to the wall-normal distribution in the WNLOP solution (see Fig. \ref{Fig:profiles_combined}) rather than the amplitude of the oblique wave, and the comparison with random initial disturbance used as the planar wave is investigated in Appendix \ref{sec:Random noises application}. 

To demonstrate the robustness of the predictive accuracy in the proposed formulation, we present the cases of $\bm{k}_2 = (1, 0), (0.25, 0)$ at $\delta_2 = 0.004$. In the case of $\bm{k}_2 = (1, 0)$ at $t_f = 50$, the P--C(5) iteration did not converge, and we used $t_f = 30$ instead of $t_f = 50$. The RMS error is $6.6 \times 10^{-5}$ at $\bm{k}_2 = (0.25, 0)$ and $3.4 \times 10^{-4}$ at $\bm{k}_2 = (1.0, 0)$, demonstrating the high predictive accuracy of the proposed methods with different wavenumber components. The decrease of predictive accuracy at $\bm{k}_2 = (1, 0)$ relates to the high initial amplification in the planar wave $F[1,0](t_f = 15) / F[1,0](t=0) \approx 3.8$ from the DNS results, which is comparable to the LOP prediction in Fig. \ref{Fig:eigenvalues} (b) of $\sqrt{G}(t_f = 15) = 4.48$ and the peak time also corresponds to the 2D LOP results. The result of $\bm{k}_2 = (0.25, 0)$ shows that the nonlinear interactions among the oblique wave and the higher harmonic waves of the streak are critical in the transition process, where a group of wavenumbers at $\bm{k} = (0.5, 1), (0.5, 2), (0.5, 3)$ is also observed for $t > 150$ as in the cases of $\bm{k}_2 = (0.5, 0)$. These results demonstrate the robustness in predictive accuracy of the proposed formulation with different wavenumber components. 

\begin{figure}
  \centering
  \begin{minipage}{0.495\textwidth}
    \centering
    \includegraphics[width=6cm]{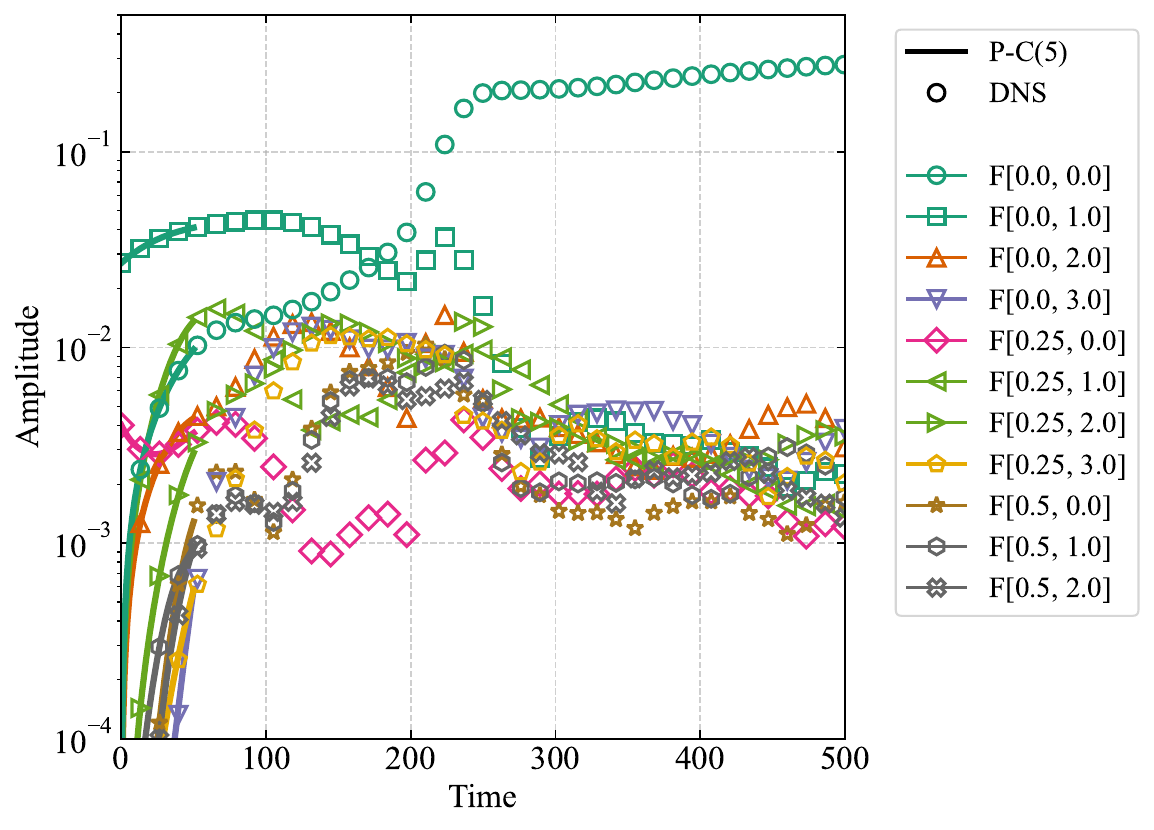}
    \subcaption{$\bm{k}_2 = (0.25, 0), t_f = 50$}
  \end{minipage} 
  \begin{minipage}{0.495\textwidth}
    \centering
    \includegraphics[width=6cm]{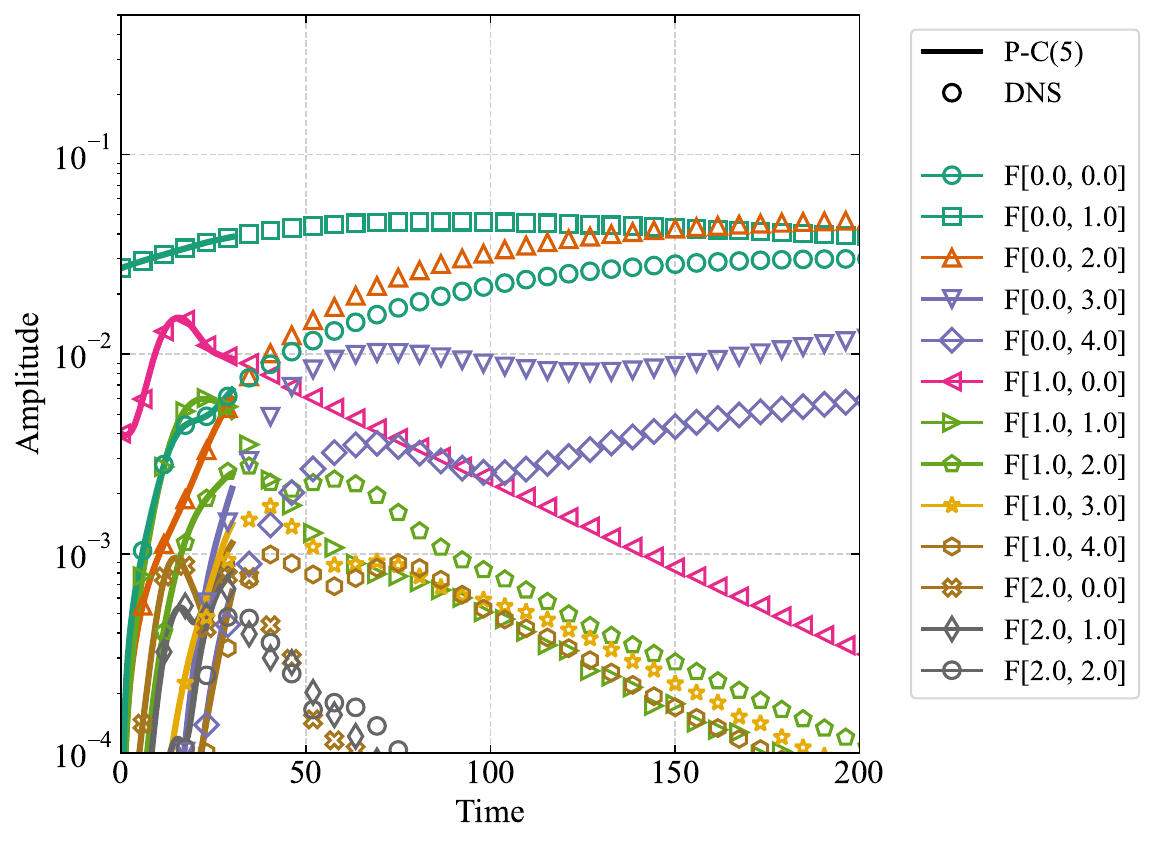}
    \subcaption{$\bm{k}_2 = (1.0, 0), t_f = 30$}
  \end{minipage} 
  \caption{
  Comparison of the P--C(5) and DNS calculations in the same manner as in Fig. \ref{Fig:pic_freq_d004_T0050_N0200_M0200_kx050} at $\delta_2=0.004$ for different wavenumbers (a) $\bm{k}_2 = (0.25, 0)$, (b) $\bm{k}_2 = (1.0, 0)$. Note that the horizontal axes have different scales. 
  }
  \label{Fig:pic_freq_d004_T0050_N0200_M0400_kx025}
\end{figure}

\begin{figure}
  \centering
  \begin{minipage}{0.495\textwidth}
    \centering
    \includegraphics[width=6cm]{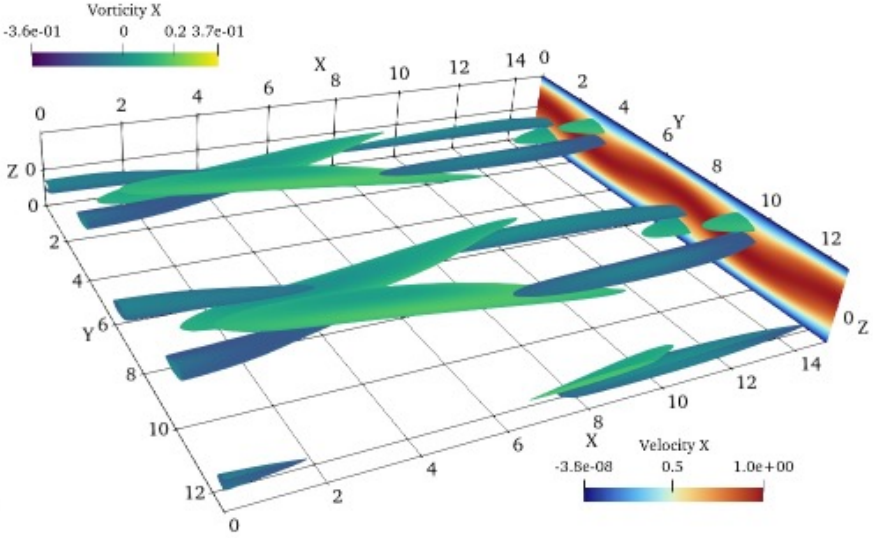}
    \subcaption{$\delta_2 = 0.002, t=53.4$}
  \end{minipage} 
  \begin{minipage}{0.495\textwidth}
    \centering
    \includegraphics[width=6cm]{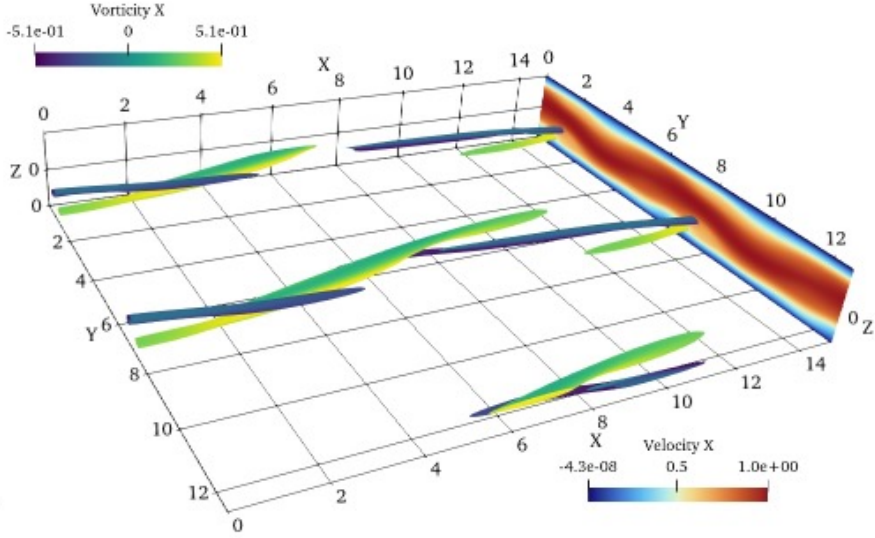}
    \subcaption{$\delta_2 = 0.004, t=53.4$}
  \end{minipage} \\
  \begin{minipage}{0.495\textwidth}
    \centering
    \includegraphics[width=6cm]{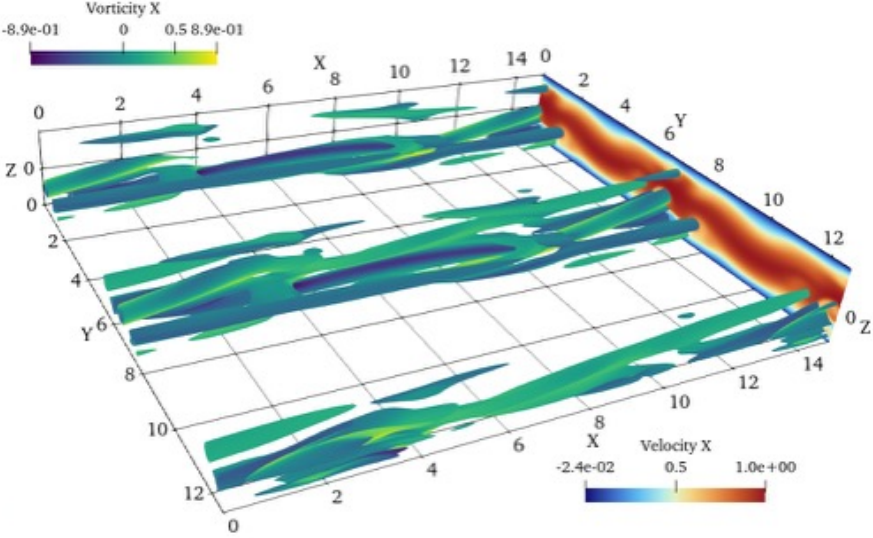}
    \subcaption{$\delta_2 = 0.004, t=85.5$}
  \end{minipage} 
  \begin{minipage}{0.495\textwidth}
    \centering
    \includegraphics[width=6cm]{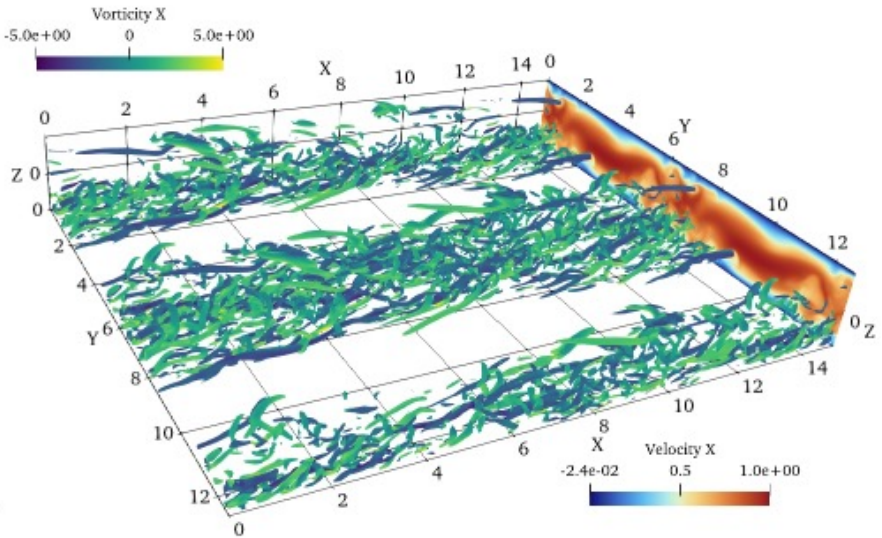}
    \subcaption{$\delta_2 = 0.004, t=117.6$}
  \end{minipage}
  \caption{
Iso-surfaces of the $Q$ criterion colored by streamwise vorticity, where the streamwise direction is restricted to $x \in [0, 15]$ of the entire computational domain ($Q$-criterion: (a) $Q=5 \times 10^{-4}$, (b) and (c) $Q=0.01$, (d) $Q=1.0$). The contour shows the streamwise velocity distribution at $x = 15$. 
  }
  \label{Fig:pic_vor_T53p44}
\end{figure}

Figure \ref{Fig:pic_vor_T53p44} visualizes the vortex structures in both cases of $\delta_2 = 0.002$ and $0.004$, represented by iso-surfaces of the second invariant of the velocity gradient tensor (the $Q$-criterion). These vortex structures are characterized by the alternating streamwise vortices inclined with respect to the spanwise direction clustered near the low-speed streaks after the initial formation of oblique waves (the same structure is also confirmed at $\delta_2 = 0.003$). In Fig. \ref{Fig:pic_vor_T53p44} (a), these vortex structures are formed on both top and bottom boundary walls, which originate from the symmetric spatial distribution of the identified WNLOP structure, as presented in Fig. \ref{Fig:profiles_combined}. This structure is similar to the NLOP structure reported by Farano et al.~\cite{farano2016subcritical} (weakly NLOP in their classification in Fig. 3 therein), despite the difference in identification methodologies between the full NLOP and the present WNLOP analysis. This alternating vortex structure induces smaller-scale vortices, as shown in Fig. \ref{Fig:pic_vor_T53p44} (c), which corresponds to the emergence of a group of modes at $\bm{k} = (0.5, 3), (1, 2), (1, 1)$ in Fig. \ref{Fig:pic_freq_d004_T0050_N0200_M0200_kx050} (c). The subsequent nonlinear interactions induce the excitation of the small and isotropic vortices near the low-speed streak on the bottom boundary wall, and this turbulent regime spreads throughout the entire domain with further time evolution. 

\begin{figure}
  \centering
  \begin{minipage}{0.49\textwidth}
    \centering
    \includegraphics[width=4.5cm]{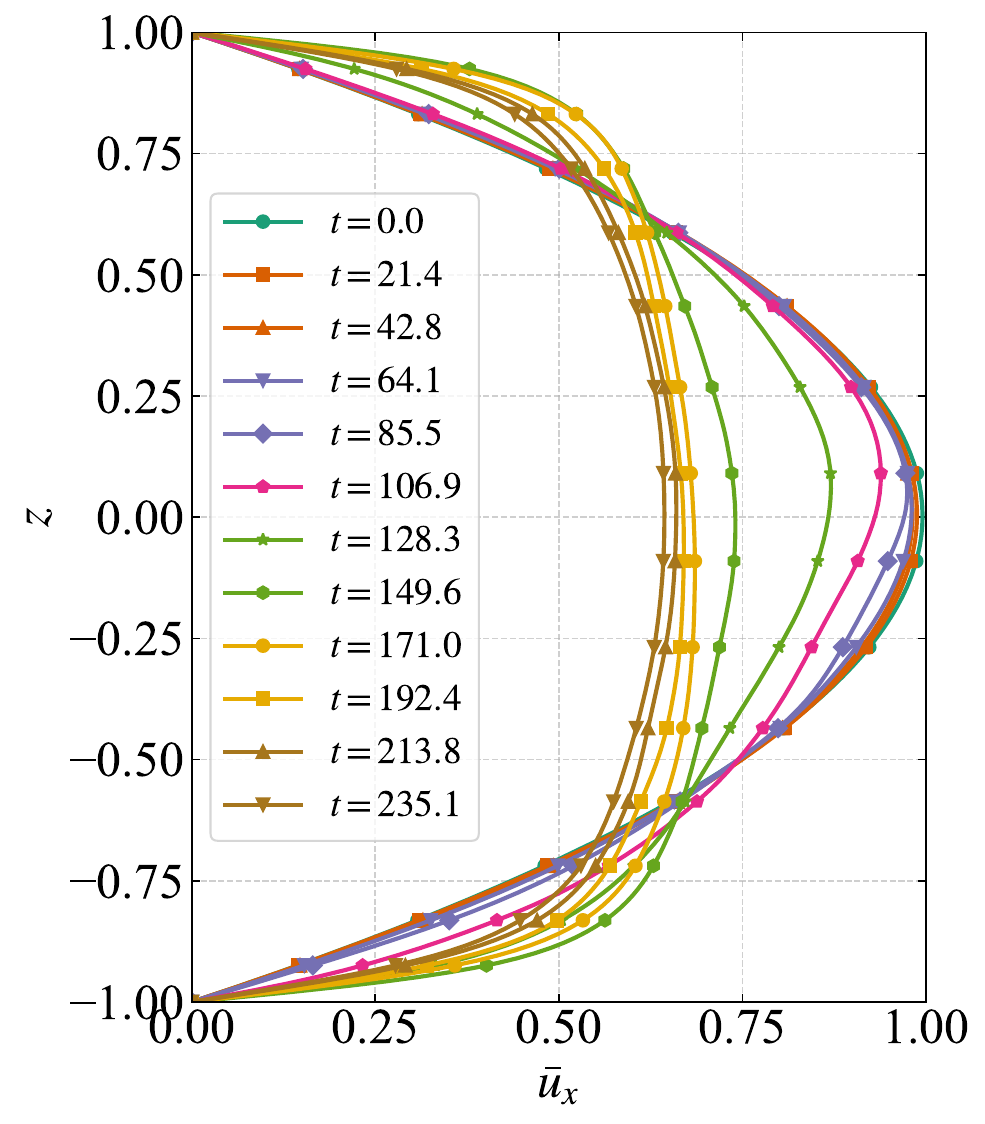}
    \subcaption{$\bar{u}_x$}
  \end{minipage}
  \begin{minipage}{0.49\textwidth}
    \centering
    \includegraphics[width=4.5cm]{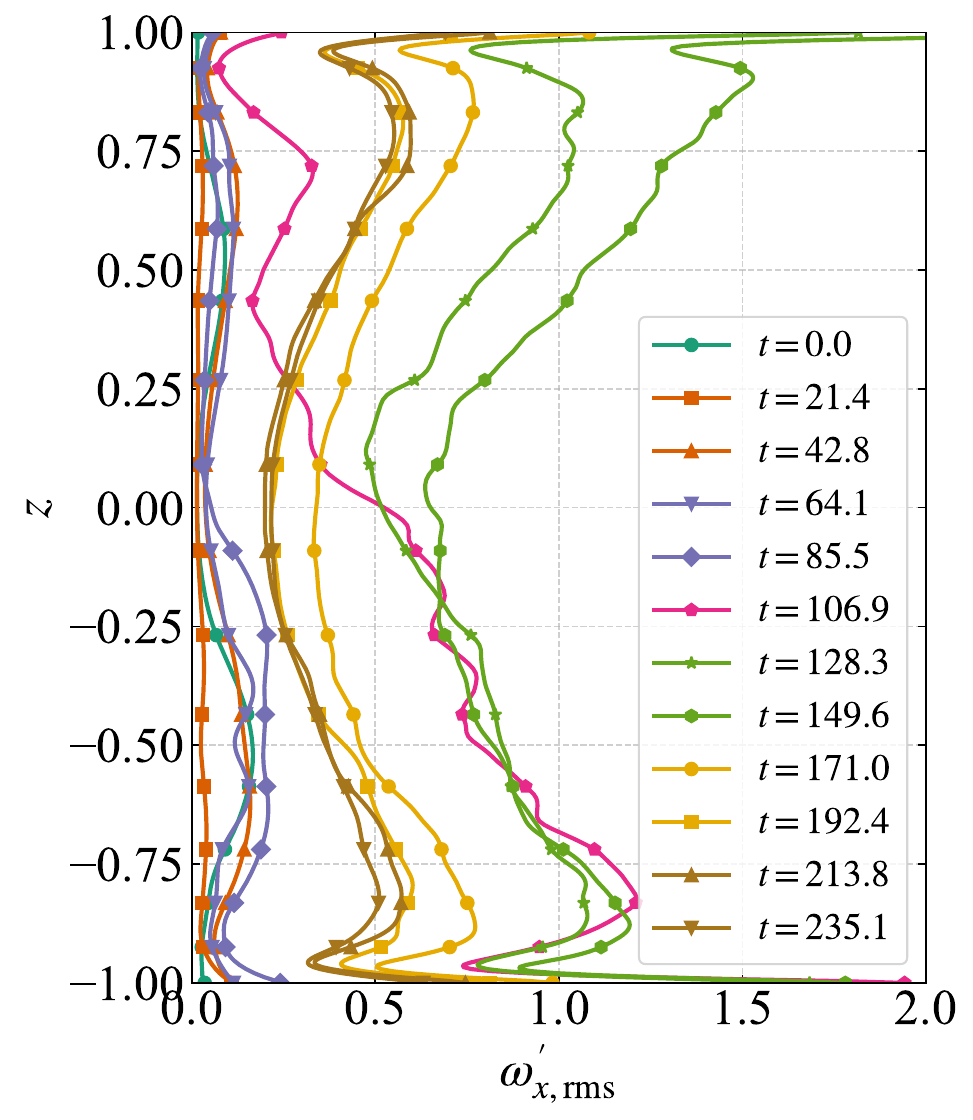}
    \subcaption{$\omega_{x, \mathrm{rms}}'$}
  \end{minipage} \\
  \begin{minipage}{0.49\textwidth}
    \centering
    \includegraphics[width=4.5cm]{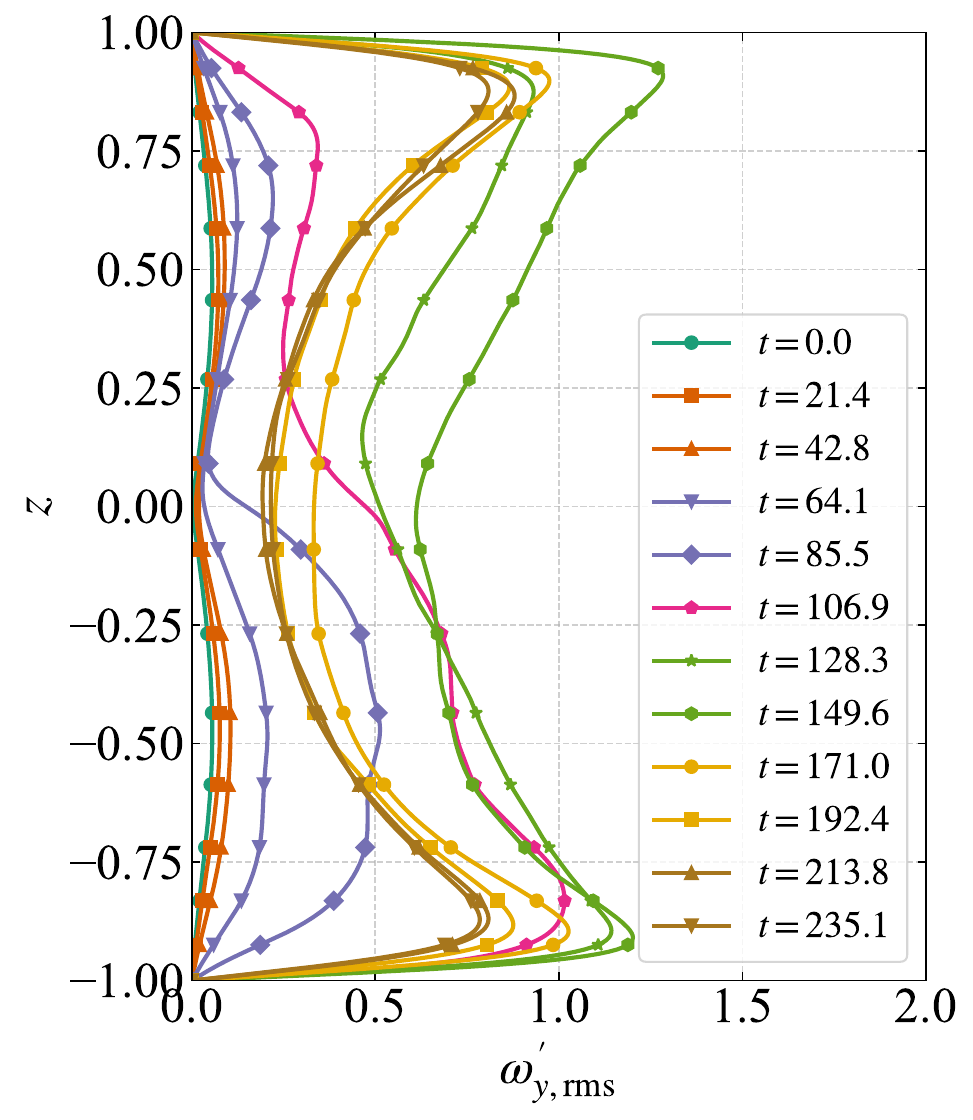}
    \subcaption{$\omega_{y, \mathrm{rms}}'$}
  \end{minipage}
  \begin{minipage}{0.49\textwidth}
    \centering
    \includegraphics[width=4.5cm]{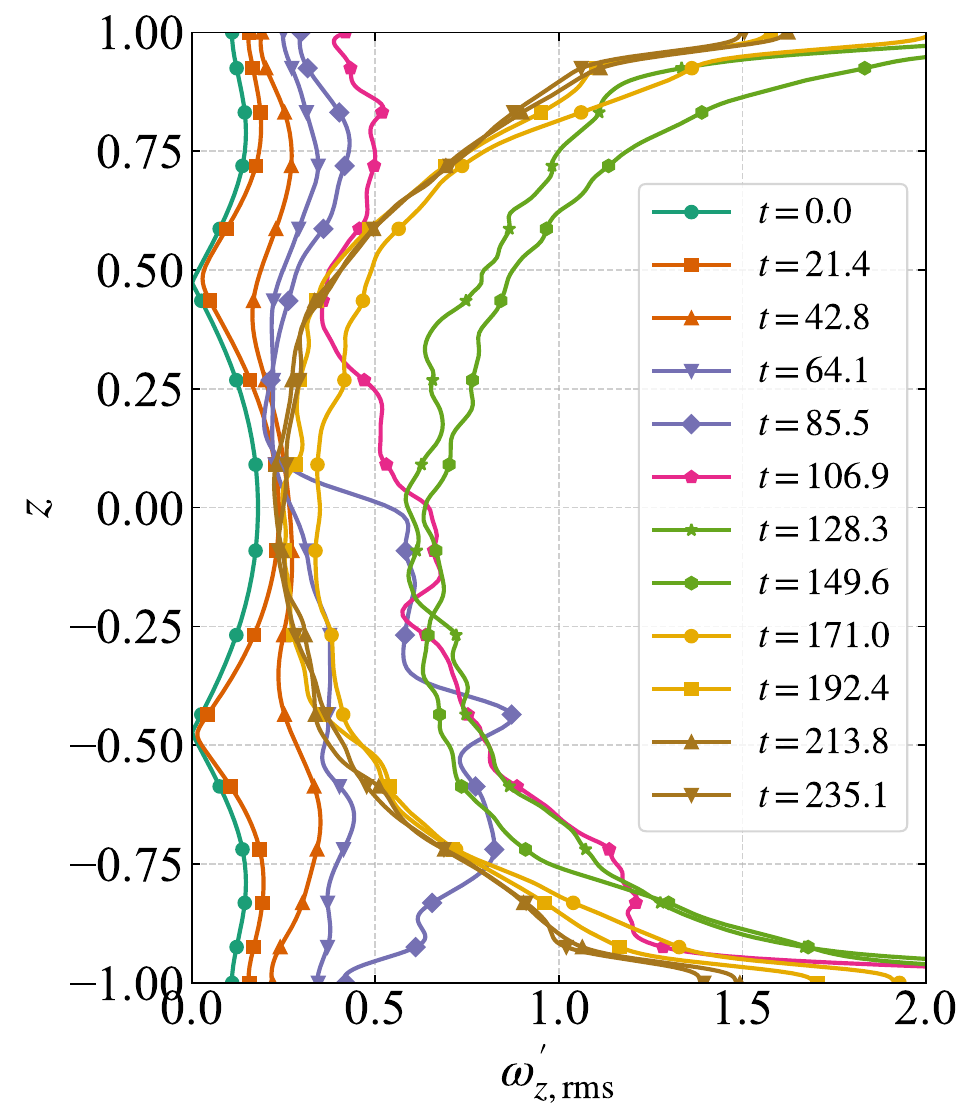}
    \subcaption{$\omega_{z, \mathrm{rms}}'$}
  \end{minipage}
  \caption{
Wall-normal distribution of (a) mean streamwise velocity and (b)-(d) the RMS of disturbance vorticity components at each snapshot. The WNLOP in the initial condition is calculated at $t_f = 50, \delta_2 = 0.004$. 
  }
  \label{Fig:omegauux_M0400_T0050}
\end{figure}

Figure \ref{Fig:omegauux_M0400_T0050} shows the wall-normal distribution of mean streamwise velocity $\bar{u}_x$ and the root mean square (RMS) of disturbance vorticity distributions $\omega_{x}', \omega_y', \omega_z'$ at each snapshot. From this figure, the mean velocity distribution transitions from the initial parabolic profile of plane Poiseuille flow to the turbulent flat distribution during time evolution. At the initial condition, the vorticity distribution is a superposition of the LOP mode in $\hat{\bm{q}}^1$ and WNLOP as a plane wave in $\hat{\bm{q}}^2$, where the asymmetric distribution of $\omega_x'$ is attributed to the WNLOP, as discussed in Fig. \ref{Fig:profiles_combined}. During the time evolution, this large amplitude of $\omega_x'$ initially decays and the amplitude of $\omega_x', \omega_y'$ is enhanced while maintaining an asymmetric distribution (see $t=64.1$ distribution), which is explained by the existence of streamwise vortices inclined in the spanwise direction, as presented in Fig. \ref{Fig:pic_vor_T53p44}. This evolution process agrees with the well-known secondary instability process of the streak reported by Cossu et al.~\cite{cossu2011secondary}. The strong nonlinear interaction amplifies the spanwise vorticity and leads to the formation of near-wall structures. 

\section{Conclusion}

In this study, a new theoretical framework of weakly nonlinear optimal perturbation (WNLOP) analysis for polychromatic waves based on Picard--Chebyshev (P--C) iteration is proposed to investigate the early nonlinear interaction leading to laminar-to-turbulent transition. The transient growth of the finite-amplitude disturbances was considered by tracking all the combinations of nonlinear interactions initiated from a few discrete wavenumber components with an increase of the iteration in the P--C(n) method, where the evolution of each wavenumber component was tracked as coupled amplitude equations. The effectiveness of the proposed formulation was carefully validated against DNS calculations, where the numerical results showed agreement with the amplitude expansion of the weakly nonlinear theory at iteration $n=3$ and approached the DNS results with an increase of $n$ in the weakly nonlinear regime. The proposed methodology has the advantage of systematically increasing accuracy beyond the limitations of conventional weakly nonlinear theory, bridging the theoretical gap between LOP and NLOP. However, the numerical results exhibited a spurious amplification of high wavenumber components in the strongly nonlinear regime, since the predictive accuracy strictly depends on the convergence of the Picard iteration, indicating the limitation of the present formulation. 

The contribution of the present formulation to the study of laminar-to-turbulent transition is to identify the existence of a key optimal structure formed from the early nonlinear interaction within the weakly nonlinear regime. The identification of the optimal planar wave that causes the largest nonlinear interactions in the presence of the streak is formulated and solved as an optimization problem in the WNLOP analysis. The optimal planar wave is composed only of spanwise velocity components, generating streamwise disturbance vorticity. The nonlinear interaction between the streak and the planar wave generates an oblique wave, and subsequently excites harmonic waves via nonlinear interactions between the streak and the oblique wave. This structure exhibits the characteristic of alternating streamwise vortices inclined in the spanwise direction, which triggers the secondary instability leading to the streak breakdown. The tracking of wavenumber components of the DNS calculations initiated under different WNLOP conditions in Fig. \ref{Fig:pic_freq_d004_T0050_N0200_M0200_kx050} shows that the amplitude of these excited wavenumber components determines the laminar-to-turbulent transition, depending on whether the amplitude is sufficient to trigger other wavenumber components in the later stages of evolution. Overall, while its predictive accuracy is restricted in the strong nonlinear regime, the present formulation provides a theoretical prediction of weakly nonlinear interactions to trigger the breakdown to complement DNS calculations, offering potential applicability to a wide class of breakdown scenarios in compressible flow between parallel plates.

\bmhead{Acknowledgements}

This study is supported by JST CREST JPMJCR24Q6.

\section*{Declarations}

The authors declare no conflict of interests. The data to support the findings of this study is available from the corresponding author upon reasonable request.

\begin{appendices}

\section{Nonlinear perturbation equations}
\label{sec:Nonlinear perturbation equations}

The pressure fluctuation are represented as $\bar{p} = \bar{\rho} \bar{T},  p' = \bar{\rho} T' + \rho' \bar{T}, p'' = \rho' T'$. Viscous dissipation term $\chi = \partial_j u_i \tau_{ij}$ and thermal diffusion term $\Theta = \chi \partial_j T$ are expanded as 
\begin{equation}
\chi = \bar{\chi} + \chi' + \chi'' + \chi''' + o(|q'|^4), \   \Theta = \bar{\Theta} + \Theta' + \Theta'' + \Theta''' + o(|q'|^4). 
\end{equation}
Taking perturbation expansions, the right-hand-side of linearized compressible N--S equations can be found as follows: 
\begin{align}
\tilde{L}_{\rho} &= - \partial_j (\bar{\rho} u_j' + \rho' \bar{u}_j), \\
\tilde{L}_{u_i} &= - (\bar{u}_j \partial_j u_i' + u_j' \partial_j \bar{u}_i) - f' \bar{u}_j \partial_j \bar{u}_i - \bar{C}  \partial_i p' +  \bar{D} \partial_j \tau_{ij}', \\
\tilde{L}_{T} &= - (u_j \partial_j T)' - f' \bar{u}_j \partial_j \bar{T} - \tilde{\gamma} ( (T \partial_j u_j)' + f' \bar{T} \partial_j \bar{u}_j) + \bar{A} \partial_j \Theta' + \bar{B} \chi', 
\end{align}
where we introduced $f' = \rho'/ \bar{\rho}, \bar{A} = A/\bar{\rho}, \bar{B} = B/\bar{\rho}, \bar{C} = 1/(\bar{\rho} \gamma \mathrm{M}^2), \bar{D} = 1/(\bar{\rho} \mathrm{Re})$ and $(fg)' = \bar{f} g' + f' \bar{g}$ for any variables $f, g$. The second-order expansions are 
\begin{align}
\tilde{N}_{1 \rho} &= - \partial_j (\rho' u_j'), \\
\tilde{N}_{1 u_i} &= - u_j' \partial_j u_i' - \bar{C} ( 
f'^2 \partial_i \bar{p} - f' \partial_i p' + \partial_i p'' ) + \bar{D} F_2 [f', \partial_j \tau_{ij}] , \\
\tilde{N}_{1 T} &= - u_j' \partial_j T' - \tilde{\gamma} T' \partial_j u_j' + \bar{A} F_2 [f', \Theta] + \bar{B} F_2 [f', \chi], 
\end{align}
where $F_2 [f', g] := f'^2 g - f' g' + g''$ for any $g$ is defined. The third-order expansions are 
\begin{align}
\tilde{N}_{2 \rho} &= 0, \\
\tilde{N}_{2 u_i} &= -\bar{C} (- f'^3 \partial_i \bar{p} + f'^2 \partial_i p' - f' \partial_i p'' ) +  \bar{D} F_3 [f', \partial_j \tau_{ij}], \\
\tilde{N}_{2 T} &= \bar{A} F_3 [f', \Theta] + \bar{B} F_3 [f', \chi] , 
\end{align}
where $F_3 [f', g] := -f'^3 \bar{g} + f'^2 g' - f' g'' + g'''$ for any $g$ is defined. Taking average of $\tilde{L}, \tilde{N}_1, \tilde{N}_2$ swapping the arguments to obtain a symmetric $L, N_1, N_2$, respectively. 

\section{Random noises application}
\label{sec:Random noises application}

\begin{figure}
  \centering
  \begin{minipage}{0.495\textwidth}
    \centering
    \includegraphics[width=6cm]{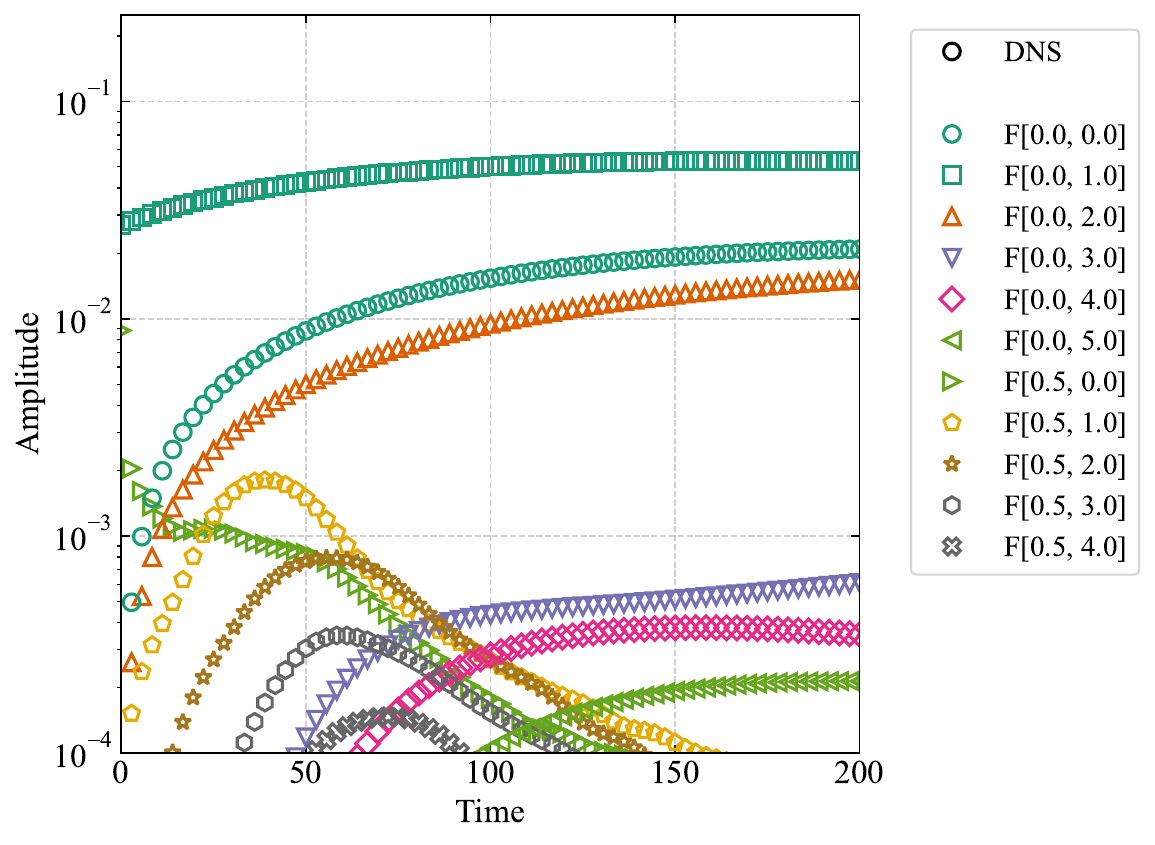}
    \subcaption{$\delta_2 = 0.01$}
  \end{minipage} 
  \begin{minipage}{0.495\textwidth}
    \centering
    \includegraphics[width=6cm]{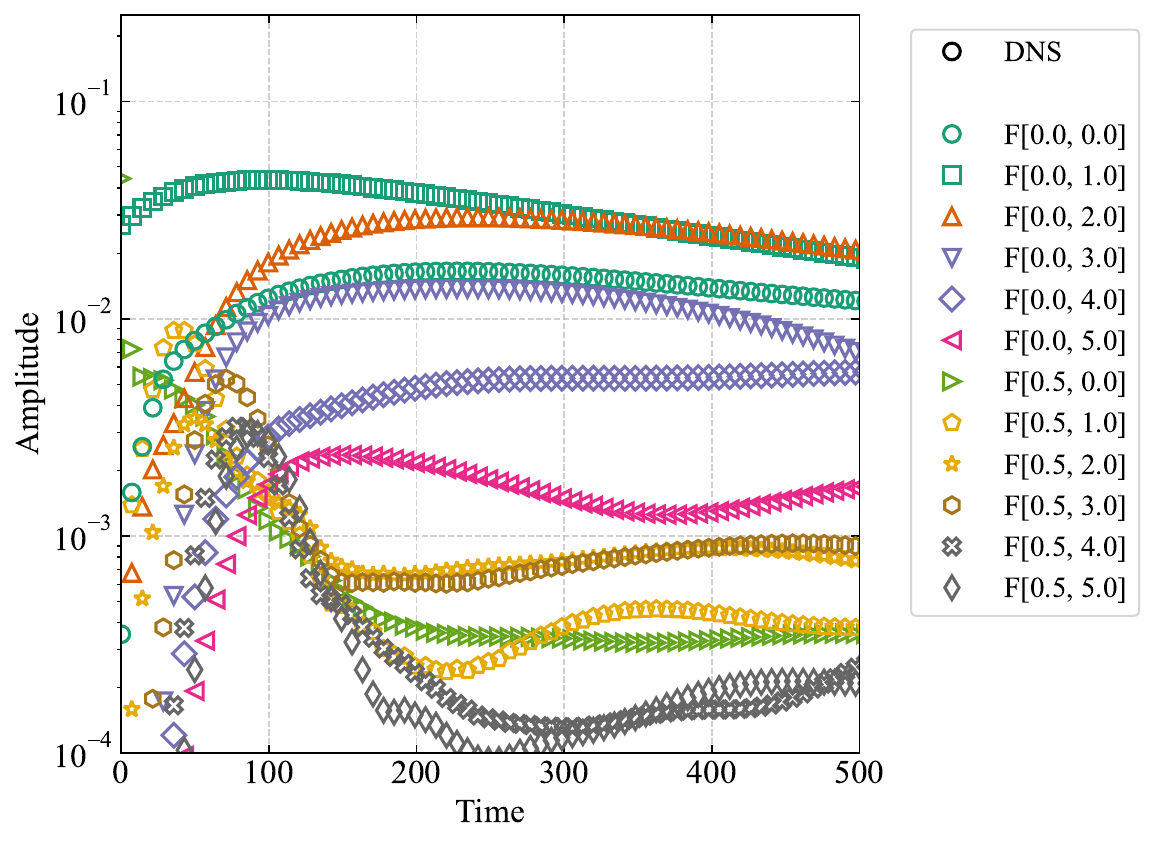}
    \subcaption{$\delta_2 = 0.05$}
  \end{minipage} 
  \caption{
  Evolution of wavenumber components in DNS calculations initiated from random velocity disturbances at $\bm{k}_2=(0.5, 0)$ instead of WNLOP solution in Fig. \ref{Fig:pic_freq_d004_T0050_N0200_M0200_kx050}. The threshold amplitudes for visualization are $F[k_x, k_y] > 10^{-4}$ at $\delta_2 = 0.01$ and $10^{-3}$ at $\delta_2 = 0.05$. Note that the horizontal axes have different scales. 
  }
  \label{Fig:pic_freq_d004_T0050_N0200_M1000_kx050_rand}
\end{figure}

In this appendix, we present numerical simulation results using random noise instead of WNLOP solutions at the wavenumber $\bm{k}_2 = (0.5, 0)$ as in Fig. \ref{Fig:pic_freq_d004_T0050_N0200_M1000_kx050_rand} to qualitatively compare the effect of the optimization solutions on the time-evolution. The random velocity noise with initial disturbance amplitudes of $\delta_2 = 0.01, 0.05$ is applied and the time-evolution is observed. The random noises initially decay due to viscous dissipation, but the case at $\delta_2 = 0.05$ retains the amplitude of the planar wave $F[0.5, 0] \approx 5.3 \times 10^{-3}$ at $t=20$ and that of an oblique wave ($\bm{k}_2 = (0.5, 1)$) with an amplitude of same-order as in the cases of $\delta_2 = 0.003, 0.004$ in Fig. \ref{Fig:pic_freq_d004_T0050_N0200_M0200_kx050}. Despite the same-order amplitude of the early nonlinear interaction processes, the streak breakdown is significantly delayed compared to the case of WNLOP solution, demonstrating the effectiveness of the proposed formulation. 

\end{appendices}


\bibliography{ref_2024_comp_poiseuille}

\end{document}